\documentclass[useAMS]{mn2e}
\usepackage{psfig}

\begin{document}

\title[The VLT-UVES survey for H$_2$ in DLA systems]{The VLT-UVES survey
for molecular hydrogen in high-redshift
damped Lyman-$\alpha$ systems\thanks{Based on observations carried out at
the European Southern Observatory (ESO), under visitor mode progs. ID
65.O-0063, 66.A-0624, 67.A-0078, 68.A-0106 and 68.A-0600, with the
UVES echelle spectrograph installed at the ESO Very Large Telescope
(VLT), unit Kueyen, on mount Paranal in Chile.}
}

\author[C\'edric Ledoux, Patrick Petitjean, \& R. Srianand]{C\'edric
Ledoux$^1$, Patrick Petitjean$^{2,3}$, \& R. Srianand$^4$\\
   $^1$ European Southern Observatory, Alonso de C\'ordova 3107,
        Casilla 19001, Vitacura, Santiago, Chile
        -- email: cledoux@eso.org\\
   $^2$ Institut d'Astrophysique de Paris -- CNRS, 98bis Boulevard Arago,
        75014, Paris, France
        -- email: petitjean@iap.fr\\
   $^3$ LERMA, Observatoire de Paris, 61 Avenue de l'Observatoire,
        75014, Paris, France\\
   $^4$ IUCAA, Post Bag 4, Ganesh Khind, Pune 411 007, India
        -- email: anand@iucaa.ernet.in\\
}

\date{Received date / Accepted date}
\pubyear{2003} \volume{000} \pagerange{1}

\maketitle \label{firstpage}

\begin{abstract}

We have searched for molecular hydrogen in damped Lyman-$\alpha$ (DLA) and
sub-DLA systems at high redshift ($z_{\rm abs}>1.8$) using UVES at the VLT
down to a detection limit of typically $N($H$_2)=2\times 10^{14}$ cm$^{-2}$.
Out of the 33 systems in our sample, 8 have firm and 2 have
tentative detections of associated H$_2$ absorption lines. Considering that
3 detections were already known from past searches, molecular hydrogen
is detected in 13 to 20 percent of the newly-surveyed systems. We report
new detections of molecular hydrogen at $z_{\rm abs}=2.087$ and 2.595 toward,
respectively, Q\,1444$+$014 and Q\,0405$-$443, and also reanalyse the
system at $z_{\rm abs}=3.025$ toward Q\,0347$-$383.

%\par\noindent
In all of the systems, we measure metallicities relative to Solar, [X/H] (with
either X$=$Zn, or S, or Si), and depletion factors of Fe, [X/Fe],
supposedly onto dust grains, and compare the characteristics of our
sample with those of the global population of DLA systems (60 systems
in total). We find that there is a correlation between metallicity and
depletion factor in both our sample and also the global population of
DLA systems. Although H$_2$ molecules are detected in systems with [Zn/Fe] as
small as 0.3, the DLA and sub-DLA systems where H$_2$ is detected are usually
amongst those having the highest metallicities and the largest depletion
factors. In particular, H$_2$ is detected in the five systems having
the largest depletion factors. Moreover, the individual components
where H$_2$ is detected have depletion factors systematically larger
than other components in the profiles. In two different systems, one of
the H$_2$-detected components even has [Zn/Fe$]\ge 1.4$. These are the largest
depletion factors ever seen in DLA systems. All this clearly demonstrates the
presence of dust in a large fraction of the DLA systems.

%\par\noindent
The mean H$_2$ molecular fraction, $f=2N($H$_2)/[2N($H$_2)+N($H\,{\sc i}$)]$,
is generally small in DLA systems (typically $\log f<-1$) and similar to what
is observed in the Magellanic Clouds. There is no correlation between the
observed amount of H$_2$ and the H\,{\sc i} column density. In fact, two
systems where H$_2$ is detected have $\log N($H\,{\sc i}$)<20.3$ and,
therefore, are sub-DLA systems. From 58 to 75 percent of the DLA systems
have $\log f<-6$. This can be explained if the formation rate of H$_2$ onto
dust grains is reduced in those systems, probably because the gas is
warm ($T>1000$ K) and/or the ionizing flux is enhanced relative to what
is observed in our Galaxy.

\end{abstract}


\section{Introduction}

High-redshift damped Lyman-$\alpha$ (DLA) systems observed in absorption
against QSO spectra are characterized by their extremely
strong H\,{\sc i}\,$\lambda$1215 lines corresponding to large neutral
hydrogen column densities, $N($H\,{\sc i}$)\ge 2\times 10^{20}$ cm$^{-2}$.
Hydrodynamical simulations suggest that DLA systems are located inside regions
of over-densities of the order of 1000 and higher
(see e.g. Haehnelt, Steinmetz \& Rauch 2000; Gardner et al. 2001) and that DLA
systems at $z_{\rm abs}\ga 2$ occur very close (within 10-15 kpc) to the
center of L$^\star$ galaxies. This is purely speculative however and the
exact nature of DLA systems is still to be clarified. Though observational
studies of DLA systems have been pursued over two decades now, important
questions are still unanswered, such as: (i) the presence of in-situ
star-formation activity in DLA systems, (ii) the connection between
observed abundance ratios and dust content, (iii) how severe is the bias due to
dust obscuration in current DLA samples. Assessing the molecular content of DLA
systems can provide direct handle on at least some of these issues.

Formation of H$_2$ is expected on the surface of dust grains if the gas is
cool, dense and mostly neutral, and from the formation of negative hydrogen
if the gas is warm and dust-free
(see e.g. Jenkins \& Peimbert 1997; Cazaux \& Tielens 2002). As the former
process is most likely dominant in the neutral gas associated with DLA
systems, it is possible to obtain an indirect indication of the amount of dust
in DLA systems without depending on extinction and/or heavy element dust
depletion effects. Moreover, from determining the populations of
different rotational levels of H$_2$, one can constrain kinetic and rotational
excitation temperatures as well as particle densities. The
effective photo-dissociation of H$_2$ takes place in the energy range
$11.1-13.6$ eV through Lyman- and Werner-band absorption lines and, therefore,
the intensity of the local UV radiation field can be derived from the observed
molecular fraction. A direct determination of the local UV radiation
field could have important implications in bridging the link between
DLA systems and star-formation activity in high-redshift galaxies.

At H\,{\sc i} column densities as large as those measured in DLA
systems, H$_2$ molecules are conspicuous in our Galaxy: gas clouds with
$\log N($H\,{\sc i}$)>21$ usually have $\log N($H$_2)>19$
(see e.g. Savage et al. 1977; Jenkins \& Shaya 1979). Given this fact, it is
somewhat surprising that earlier searches for molecular hydrogen in DLA
systems, though not systematic, have found either small values or upper limits
on the molecular fraction of the
gas, $f=2N($H$_2)/[2N($H$_2)+N($H\,{\sc i}$)]$ (Black, Chaffee \& Foltz 1987;
Levshakov et al. 1992). For a long time, only the DLA system at
$z_{\rm abs}=2.811$ toward Q\,0528$-$250 was known to contain
molecular hydrogen (Levshakov \& Varshalovich 1985; Foltz, Chaffee \& Black 1988).
More recently, Ge \& Bechtold (1999) searched for H$_2$ molecules in a sample
of eight DLA systems using the MMT moderate-resolution spectrograph
(${\rm FWHM}=1$ \AA ). Apart from the detection of molecular hydrogen
at $z_{\rm abs}=1.973$ and 2.338 toward, respectively, Q\,0013$-$004
and Q\,1232$+$082 (Ge \& Bechtold 1997; Ge, Bechtold \& Kulkarni 2001), they
measured in the other systems upper limits on $f$ in the range
$10^{-6}-10^{-4}$.

The molecular hydrogen content of the above three systems, where H$_2$
was detected at intermediate spectral resolution, has been
reexamined systematically using high spectral resolution
data (${\rm FWHM}=0.1$ \AA ), leading to:
$N($H$_2)\sim 6\times 10^{16}$ cm$^{-2}$ and $f\sim 5\times 10^{-5}$ toward
Q\,0528$-$250 (Srianand \& Petitjean 1998),
$5\times 10^{17}<N($H$_2)<10^{20}$ cm$^{-2}$ and $2\times 10^{-3}<f<0.2$
toward Q\,0013$-$004 (Petitjean, Srianand \& Ledoux 2002) and
$N($H$_2)\ge 2\times 10^{17}$ cm$^{-2}$ and $f\ge 4\times 10^{-4}$ toward
Q\,1232$+$082 (Srianand, Petitjean \& Ledoux 2000). Recent new detections are
from Levshakov et al. (2002) at $z_{\rm abs}=3.025$ toward Q\,0347$-$383
and Ledoux, Srianand \& Petitjean (2002b) at $z_{\rm abs}=1.962$ toward
Q\,0551$-$366. Two tentative detections have also been reported at
$z_{\rm abs}=2.374$ toward Q\,0841$+$129 (Petitjean, Srianand \& Ledoux 2000)
and at $z_{\rm abs}=3.390$ toward Q\,0000$-$263 (Levshakov et al. 2000)
based, however, on the identification of two weak features located in
the Lyman-$\alpha$ forest.

This paper describes the high spectral resolution survey of high-redshift DLA
systems, focused on the search for molecular hydrogen, that we have carried
out at the ESO VLT during the first two years of operations of the
UVES echelle spectrograph. We present here our final results from
the observations of a total sample of 24 DLA ($\log N($H\,{\sc i}$)\ge 20.3$)
and 9 sub-DLA ($19.4<\log N($H\,{\sc i}$)<20.3$) systems, including two new
H$_2$ detections. Preliminary results from the observations of a sample of
11 systems were presented in Petitjean et al. (2000). Sect.~\ref{obs} in
the present paper describes the observations, the data and the
absorber sample, Sect.~\ref{ind} gives details on three DLA/sub-DLA systems
where H$_2$ is detected, and in both Sects.~\ref{sam} and \ref{res}
general results are discussed from the study of the whole sample. We
conclude in Sect.~\ref{con}.

\begin{table*}
\caption{Atomic data}
\begin{tabular}{llllllll}
\hline
\hline
Transition & $\lambda _{\rm vac}$ & $f$ & Ref. $^2$ & Transition & $\lambda _{\rm vac}$ & $f$ & Ref. $^2$ \\
           & (\AA )               &     &           &            & (\AA )               &     &           \\
\hline
H\,{\sc i}\,$\lambda$1025                 & 1025.7223       & 0.07912  & a & Si\,{\sc ii}\,$\lambda$1808               & 1808.0129       & 0.00208  & f \\
H\,{\sc i}\,$\lambda$1215                 & 1215.6701       & 0.4164   & a & P\,{\sc ii}\,$\lambda$963                 &  963.801        & 1.458    & a \\
C\,{\sc i}\,$\lambda$1328                 & 1328.8333       & 0.0630   & b & P\,{\sc ii}\,$\lambda$1152                & 1152.8180       & 0.236    & a \\
C\,{\sc i}\,$^\star\lambda$1329.08        & 1329.0853       & 0.0213   & b & S\,{\sc ii}\,$\lambda$1250                & 1250.584        & 0.00545  & a \\
C\,{\sc i}\,$^\star\lambda$1329.10        & 1329.1004       & 0.0260   & b & S\,{\sc ii}\,$\lambda$1253                & 1253.811        & 0.0109   & a \\
C\,{\sc i}\,$^\star\lambda$1329.12        & 1329.1233       & 0.0160   & b & S\,{\sc ii}\,$\lambda$1259                & 1259.519        & 0.0162   & a \\
C\,{\sc i}\,$\lambda$1560                 & 1560.3092       & 0.0719   & b & Ar\,{\sc i}\,$\lambda$1048                & 1048.2199       & 0.257    & g \\
C\,{\sc i}\,$^\star\lambda$1560.6         & 1560.6822       & 0.0539   & b & Ti\,{\sc ii}\,$\lambda$1910.6             & 1910.609        & 0.104    & h \\
C\,{\sc i}\,$^\star\lambda$1560.7         & 1560.7090       & 0.0180   & b & Ti\,{\sc ii}\,$\lambda$1910.9             & 1910.938        & 0.098    & h \\
C\,{\sc i}\,$\lambda$1656                 & 1656.9283       & 0.139    & b & Cr\,{\sc ii}\,$\lambda$2056               & 2056.2569       & 0.105    & i \\
C\,{\sc i}\,$^\star\lambda$1656           & 1656.2672       & 0.0589   & b & Cr\,{\sc ii}\,$\lambda$2062               & 2062.2361       & 0.0780   & i \\
C\,{\sc i}\,$^\star\lambda$1657.3         & 1657.3792       & 0.0356   & b & Cr\,{\sc ii}\,$\lambda$2066               & 2066.1640       & 0.0515   & i \\
C\,{\sc i}\,$^\star\lambda$1657.9         & 1657.9068       & 0.0473   & b & Fe\,{\sc ii}\,$\lambda$1096               & 1096.8769       & 0.0324   & j \\
C\,{\sc i}\,$^{\star\star }\lambda$1657   & 1657.0082       & 0.104    & b & Fe\,{\sc ii}\,$\lambda$1121               & 1121.9749       & 0.0202   & j \\
C\,{\sc ii}\,$^\star\lambda$1037          & 1037.0182       & 0.123    & a & Fe\,{\sc ii}\,$\lambda$1125               & 1125.4478       & 0.0163   & j \\
C\,{\sc ii}\,$^\star\lambda$1335.6        & 1335.6627       & 0.0128   & a & Fe\,{\sc ii}\,$\lambda$1143               & 1143.2260       & 0.0177   & j \\
C\,{\sc ii}\,$^\star\lambda$1335.7        & 1335.7077       & 0.115    & a & Fe\,{\sc ii}\,$\lambda$1144               & 1144.9379       & 0.106    & j \\
N\,{\sc i}\,$\lambda$1134.1               & 1134.1653       & 0.0134   & a & Fe\,{\sc ii}\,$\lambda$1608               & 1608.4509       & 0.0585   & j \\
N\,{\sc i}\,$\lambda$1134.4               & 1134.4149       & 0.0268   & a & Fe\,{\sc ii}\,$\lambda$1611               & 1611.2005       & 0.00136  & j \\
N\,{\sc i}\,$\lambda$1134.9               & 1134.9803       & 0.0402   & a & Fe\,{\sc ii}\,$\lambda$2249               & 2249.8768       & 0.00182  & j \\
N\,{\sc i}\,$\lambda$1200.2               & 1200.2233       & 0.0885   & a & Fe\,{\sc ii}\,$\lambda$2260               & 2260.7805       & 0.00244  & j \\
O\,{\sc i}\,$\lambda$950                  &  950.8846       & 0.00157  & a & Fe\,{\sc ii}\,$\lambda$2374               & 2374.4612       & 0.0313   & j \\
O\,{\sc i}\,$\lambda$974                  &  974.07(5) $^1$ & 0.00002  & a & Ni\,{\sc ii}\,$\lambda$1317               & 1317.2170       & 0.0774   & k \\
Mg\,{\sc ii}\,$\lambda$1239               & 1239.9253       & 0.000554 & c & Ni\,{\sc ii}\,$\lambda$1370               & 1370.1320       & 0.0765   & k \\
Mg\,{\sc ii}\,$\lambda$1240               & 1240.3947       & 0.000277 & c & Ni\,{\sc ii}\,$\lambda$1741               & 1741.5531       & 0.0427   & l \\
Al\,{\sc ii}\,$\lambda$1670               & 1670.7886       & 1.833    & a & Ni\,{\sc ii}\,$\lambda$1751               & 1751.9157       & 0.0277   & l \\
Si\,{\sc ii}\,$\lambda$1020               & 1020.6989       & 0.0283   & a & Zn\,{\sc ii}\,$\lambda$2026               & 2026.1371       & 0.489    & i \\
Si\,{\sc ii}\,$\lambda$1304               & 1304.3702       & 0.0894   & d & Zn\,{\sc ii}\,$\lambda$2062               & 2062.6604       & 0.256    & i \\
Si\,{\sc ii}\,$\lambda$1526               & 1526.7070       & 0.127    & e &                                           &                 &          &   \\
\hline
\end{tabular}
\label{tabosc}
\flushleft $^1$ See Subsect.~\ref{q0347}.\\
$^2$ {\sc References} for oscillator strengths:
(a)~Morton (1991);
(b)~Wiese, Fuhr \& Deters (1996);
(c)~Welty et al. (1999);
(d)~Spitzer \& Fitzpatrick (1993);
(e)~Schectman, Povolny \& Curtis (1998);
(f)~Bergeson \& Lawler (1993b);
%(z)~Bi\'emont et al. (1994);%Cl\,{\sc i}
(g)~Federman et al. (1992);
(h)~Wiese, Fedchak \& Lawler (2001);
(i)~Bergeson \& Lawler (1993a);
(j)~Howk et al. (2000);
(k)~Fedchak \& Lawler (1999);
(l)~Fedchak, Wiese \& Lawler (2000).
\end{table*}

\section{Observations and sample}\label{obs}

The Ultraviolet and Visible Echelle Spectrograph
(UVES; Dekker et al. 2000), installed at the ESO VLT 8.2-m telescope, unit
Kueyen, on Mount Paranal in Chile, was used during five visitor-mode
runs, from April 2000 to January 2002.
During each run, we surveyed DLA and sub-DLA systems with
the aim of searching for H$_2$ lines with redshifts in the range $1.8-3.4$. In
a first step, we observed the ($V<19$) background QSOs twice for 1 to 1.5 hr
over wavelength ranges covering the location of the main metal lines of
the absorption systems and possibly associated H$_2$ features in,
respectively, the Red and the Blue arms of UVES (which were used
simultaneously). For most of the targets, this resulted to a
signal-to-noise ratio in excess of 10 in the Blue. The quick-look
data reduction package available at the telescope was extensively used
to decide in real time whether H$_2$ absorption lines are present or not.
This allowed us to optimize our observing strategy: in case of detection,
additional exposures were gathered to achieve the high signal-to-noise ratio
needed for an accurate determination of the H$_2$ column densities in
different rotational levels. In addition to our observations, we also
included the few UVES archival spectra of QSOs with DLA systems for which the
H$_2$ wavelength ranges were covered.

The total QSO sample presently under consideration is made of 25 lines
of sight. Observations toward Q\,0013$-$004, Q\,0551$-$366
and Q\,1232$+$082, along the lines of sight of which H$_2$ is detected,
are described respectively in Petitjean et al. (2002), Ledoux et al. (2002b)
and Srianand et al. (2000). In addition, new data for Q\,0528$-$250
and Q\,1232$+$082 were acquired recently and the corresponding H$_2$-detected
systems should be analysed in detail in forthcoming papers. However, we give
in Sect.~\ref{sam} the measurement from UVES data of the molecular hydrogen
and metal contents of the DLA system at $z_{\rm abs}=2.811$ toward
Q\,0528$-$250.

With respect to the remaining and/or new detections of H$_2$ in our
sample, high-resolution, high signal-to-noise ratio spectra of Q\,0347$-$383
($z_{\rm em}=3.21$, $m_{\rm V}=17.8$), Q\,0405$-$443
($z_{\rm em}=3.02$, $m_{\rm B}=17.6$) and Q\,1444$+$014 ($z_{\rm em}=2.21$,
$m_{{\rm B}_{\rm J}}=18.5$) were obtained on, respectively, January 7-9, 2002,
October 20-23, 2001, and June 16, 2001. During the observations of
Q\,0405$-$443, central wavelengths 3900 and 5700 \AA\ were used
in, respectively, the Blue and the Red arms of UVES with Dichroic \#1, and
central wavelengths were adjusted to 4370 \AA\ in the Blue and 7400 \AA\ in
the Red with Dichroic \#2. Full wavelength coverage was achieved this way
from 3283 to 7320 \AA\ and from 7468 to 9302 \AA\ accounting for the
gap between the two Red-arm CCDs. During the observations
of Q\,1444$+$014 (resp. Q\,0347$-$383), central wavelengths were adjusted
to 3800 \AA\ (resp. 4300 \AA ) in the Blue and 5640 \AA\ (resp. 8000 \AA ) in
the Red, and we used Dichroic \#1 (resp. \#2). Complementary UVES data
for Q\,1444$+$014, obtained on May 29-30, 2000, with similar settings,
were retrieved from the UVES archive. For each of the 25 quasars in our
sample, the CCD pixels were binned $2\times 2$ in both arms and the slit width
was fixed to $1\arcsec$ (except for Q\,0347$-$383: $0.8\arcsec$) yielding
a spectral resolution $R=42500$ (resp. $R=52000$). The total integration times
were $9\times 4500$ s for Q\,0347$-$383, $4\times 4350$ s with Dichroic
\#1 (and $2\times 4350$ s with Dichroic \#2) for Q\,0405$-$443,
and $8\times 3600$ s for Q\,1444$+$014.

\begin{figure*}
\centerline{\hbox{
\psfig{figure=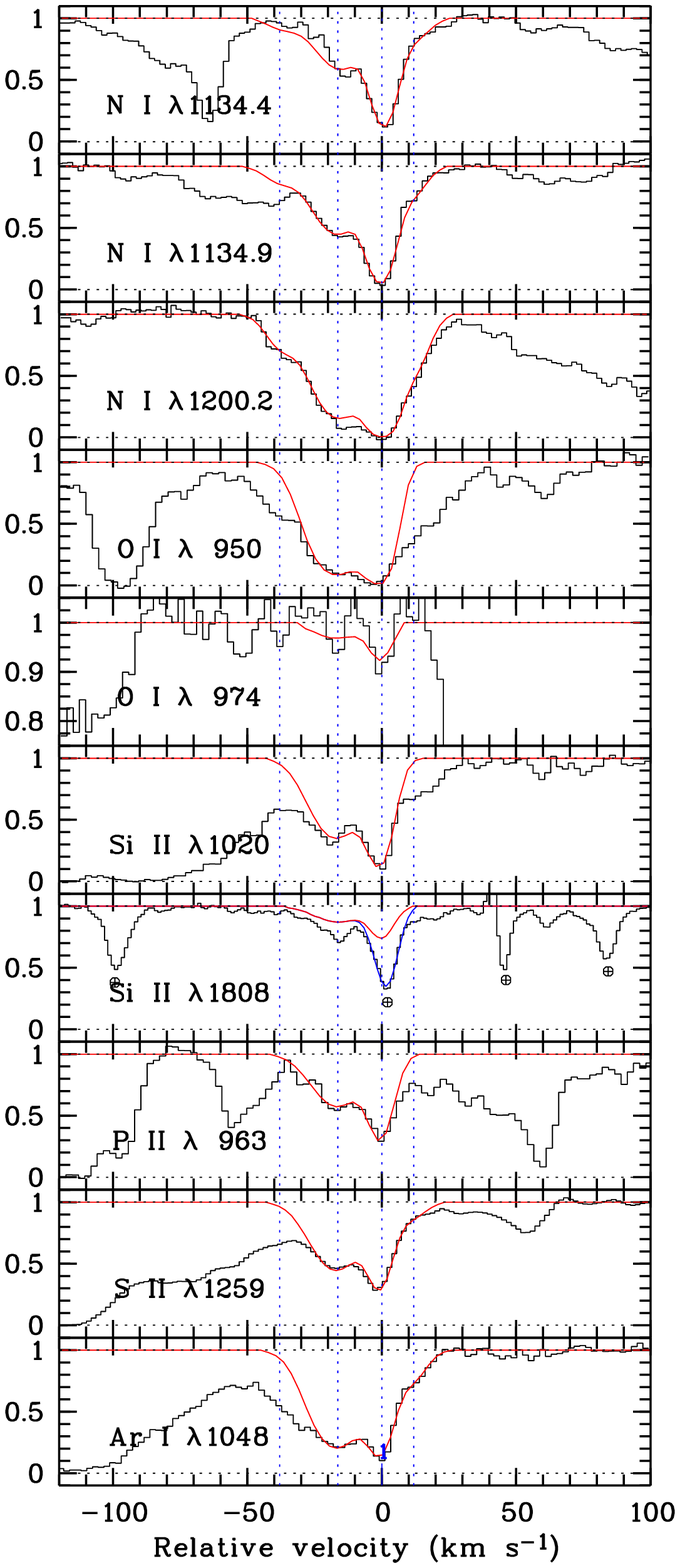,width=8.65cm,clip=,bbllx=57.pt,bblly=48.pt,bburx=370.pt,bbury=769.pt,angle=0.}\hspace{+0.3cm}\psfig{figure=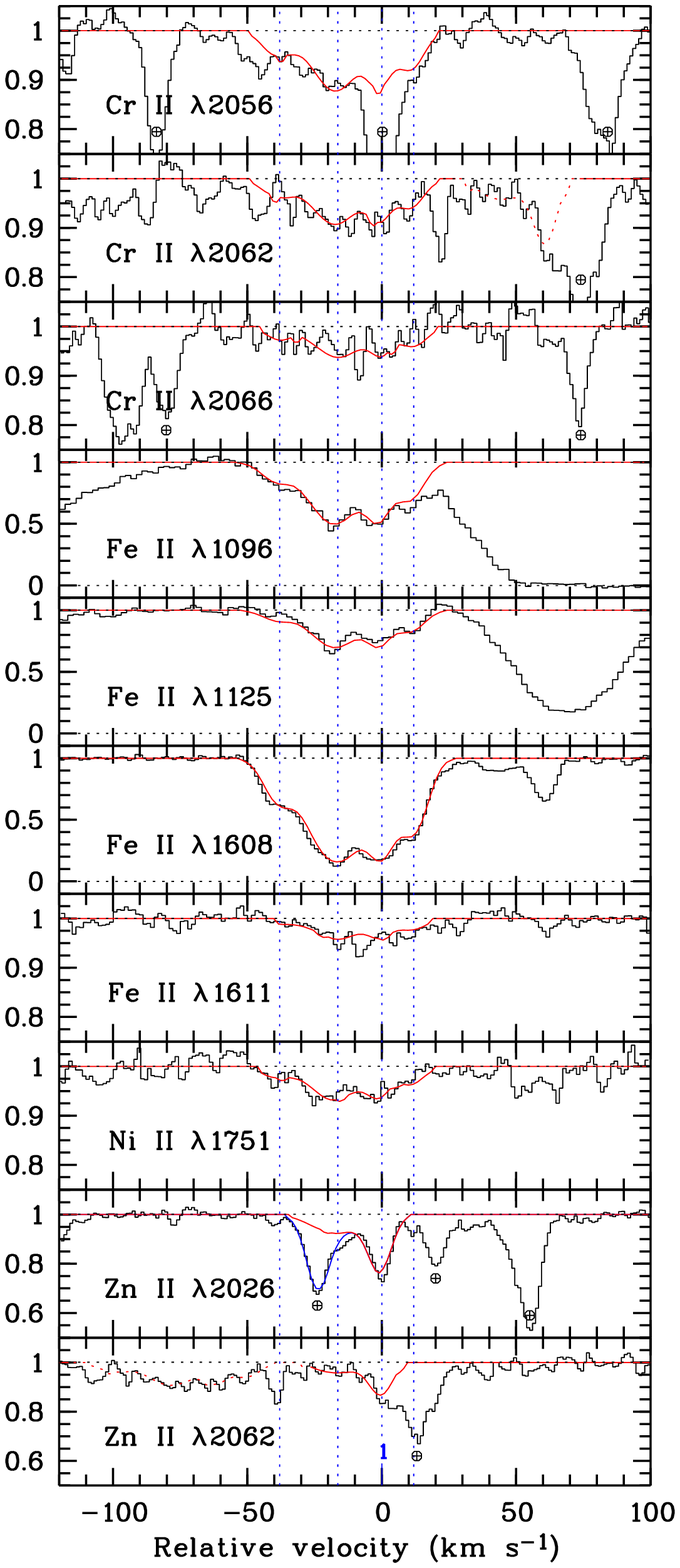,width=8.65cm,clip=,bbllx=57.pt,bblly=48.pt,bburx=370.pt,bbury=769.pt,angle=0.}}}
\caption[]{Velocity profiles of lines from low ions and neutral species in the
DLA system at $z_{\rm abs}=3.025$ toward Q\,0347$-$383. The
S\,{\sc ii}\,$\lambda$1259 profile is reproduced from
Prochaska \& Wolfe (1999). Our best-fitting model is superimposed on the
spectra with vertical lines marking the location of individual components.
The dotted parts in some of the synthetic profiles correspond to other
transitions than the ones indicated (Zn\,{\sc ii}\,$\lambda$2062 and
Cr\,{\sc ii}\,$\lambda$2062). The single component where H$_2$ is detected
is labelled with number 1. The location of telluric absorption features is
marked by the symbol $\earth$. Note that the blending with telluric lines
has been carefully taken into account (dark curves) for the fitting of
the Si\,{\sc ii}\,$\lambda$1808 and Zn\,{\sc ii}\,$\lambda$2026
profiles (light curves).}
\label{figmetals2}
\end{figure*}

\begin{figure*}
\flushleft{\vbox{
\psfig{figure=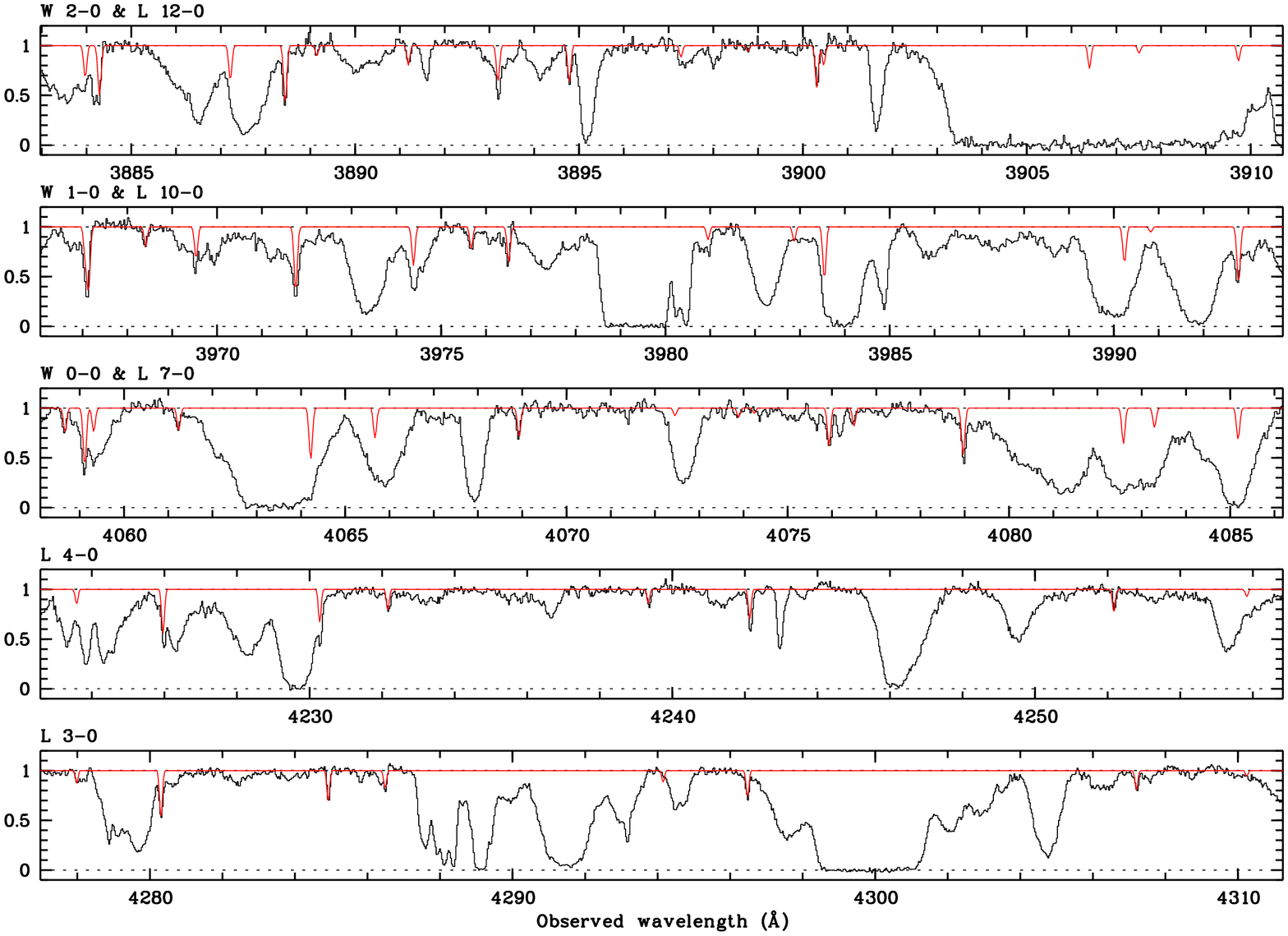,width=17.7cm,clip=,bbllx=41.pt,bblly=63.pt,bburx=555.pt,bbury=778.pt,angle=270.}}}
\caption[]{Voigt-profile fitting to the transition lines from the $J=0$, 1,
2, 3 and 4 rotational levels of the vibrational ground-state Lyman and Werner
bands of H$_2$ at $z_{\rm abs}=3.02489$ toward Q\,0347$-$383. Only a few
of the observed bands are displayed. H$_2$ is detected on this line of
sight in a single gas cloud.}
\label{figmole2}
\end{figure*}

The spectra for each QSO were reduced in the dedicated context of
{\sc MIDAS}, the
ESO data reduction system, using the UVES pipeline (Ballester et al. 2000)
in an interactive mode. The main characteristics of the pipeline are to
perform a precise inter-order background subtraction for science frames
and master flat-fields, and an optimal extraction of the object signal
rejecting cosmic ray impacts and subtracting the sky spectrum simultaneously.
The pipeline products were checked step by step. The wavelength scale of
the spectra reduced by the pipeline was then converted to vacuum-heliocentric
values and individual 1-D exposures were scaled, weighted and
combined altogether using the {\it scombine} task of the NOAO {\it onedspec}
package of {\sc IRAF}. During this process, the spectra of each QSO were rebinned to
a common wavelength step (0.05 \AA\ pix$^{-1}$ or less). In particular,
the spectra of Q\,0347$-$383, Q\,0405$-$443 and Q\,1444$+$014, where H$_2$ is
detected, were rebinned to, respectively, 0.03, 0.0471 and 0.0339 \AA\
pix$^{-1}$, yielding after combination of individual exposures
signal-to-noise ratios in, respectively, the ranges 20-50, 10-50 and 10-45 in
the Blue and 90-15, 120-20 and 90-30 in the Red.

Standard Voigt-profile fitting techniques were used to determine column
densities, in all the absorption line systems for which the H$_2$
wavelength range was observed, using the oscillator strengths compiled in
Table~\ref{tabosc} for metal ions and the oscillator strengths given
by Morton \& Dinerstein (1976) for H$_2$. In the following, we
measure metallicities relative to
Solar, [X/H$]\equiv\log [N($X$)/N($H$)]-\log [N($X$)/N($H$)]_\odot$,
with either X$=$Zn, or S, or Si, and depletion factors of
iron, [X/Fe$]\equiv\log [N($X$)/N($Fe$)]-\log [N($X$)/N($Fe$)]_\odot$,
adopting the Solar abundances from Savage \& Sembach (1996).

There has been no a priori selection of the systems. In our sample, the 24 DLA
and 9 sub-DLA absorbers are generally drawn at random from the
global population of known systems. However, we have reobserved with
UVES three DLA systems where H$_2$ was previously detected
(toward Q\,0013$-$004, Q\,0528$-$250 and Q\,1232$+$082). Finally, the
number of lines of sight along which several systems are seen is larger in
our sample than in the global population of DLA systems. One would expect this
to introduce a bias in favor of low-metallicity absorbers. However, the
characteristics of our sample and those of the global population of DLA
systems are compared in Section~\ref{sam} and they are shown to be similar.

\section{Description of individual systems with H$_2$ detection}\label{ind}

In this Section, we reconsider the case of the DLA system at
$z_{\rm abs}=3.025$ toward Q\,0347$-$383 and describe two new detections of
molecular hydrogen at, respectively, $z_{\rm abs}=2.595$ and 2.087 toward,
respectively, Q\,0405$-$443 and Q\,1444$+$014.

\subsection{Q\,0347$-$383, $z_{\rm abs}=3.025$}\label{q0347}

\begin{table}
\caption {Voigt-profile fitting results for different rotational levels of the
vibrational ground-state Lyman and Werner bands of H$_2$ toward Q\,0347$-$383}
\begin{tabular}{llll}
\hline
\hline
$z_{\rm abs}$ & Level & $\log N\pm\sigma _{\log N}$ & $b\pm\sigma _b$\\
              &       & (H$_2$)                     & (km s$^{-1}$)  \\
\hline
3.02489 & $J=0$ &     $13.25\pm 0.08$         &  $1.3\pm 0.2$    \\
        & $J=1$ &     $14.26\pm 0.06$         &  \phantom{uu..}''\\
        & $J=2$ &     $13.65\pm 0.04$         &  \phantom{uu..}''\\
        & $J=3$ &     $13.90\pm 0.04$         &  \phantom{uu..}''\\
        & $J=4$ &     $13.12\pm 0.12$         &  \phantom{uu..}''\\
        & $J=5$ & $\le 12.75$ $^{\rm a}$      &  \phantom{uu..}''\\
\hline
\end{tabular}
\label{tabmol2}
\flushleft
$^{\rm a}$ Possible blends.
\end{table}

\begin{table}
\caption {Ionic column densities in individual components of the DLA system
at $z_{\rm abs}=3.025$ toward Q\,0347$-$383}
\begin{tabular}{llll}
\hline
\hline
Ion & Transition & $\log N\pm\sigma _{\log N}$ & $b\pm\sigma _b$\\
    & lines used &                             & (km s$^{-1}$)  \\
\hline
\multicolumn{4}{l}{$z_{\rm abs}=3.02434$}\\
N\,{\sc i}   & 1134.4,1134.9,1200.2            & $13.14\pm 0.08$\phantom{$^{\rm a}$} & \phantom{0}$6.7\pm 0.8$\\
Cr\,{\sc ii} & 2056,2062,2066                  & $12.09\pm 0.14$\phantom{$^{\rm a}$} & \phantom{uuu..}''      \\
Fe\,{\sc ii} & 1096,1125,1144 $^{\rm a}$       & $13.36\pm 0.06$\phantom{$^{\rm a}$} & \phantom{uuu..}''      \\
Ni\,{\sc ii} & 1741,1751                       & $12.37\pm 0.36$\phantom{$^{\rm a}$} & \phantom{uuu..}''      \\
\hline
\multicolumn{4}{l}{$z_{\rm abs}=3.02463$}\\
N\,{\sc i}   & 1134.4,1134.9,1200.2            & $14.15\pm 0.03$\phantom{$^{\rm a}$} & $11.4\pm 0.8$          \\
O\,{\sc i}   & 950,974                         & $16.12\pm 0.12$ $^{\rm b}$          & \phantom{uuu..}''      \\
Si\,{\sc ii} & 1020,1808                       & $14.47\pm 0.03$\phantom{$^{\rm a}$} & \phantom{uuu..}''      \\
P\,{\sc ii}  & 963                             & $12.34\pm 0.19$ $^{\rm c}$          & \phantom{uuu..}''      \\
S\,{\sc ii}  & 1259 $^{\rm d}$                 & $14.50\pm 0.03$\phantom{$^{\rm a}$} & \phantom{uuu..}''      \\
Ar\,{\sc i}  & 1048                            & $13.68\pm 0.03$\phantom{$^{\rm a}$} & \phantom{uuu..}''      \\
Ti\,{\sc ii} & 1910.6,1910.9                   & $<11.86$ $^{\rm e}$                 & \phantom{uuu..}''      \\
Cr\,{\sc ii} & 2056,2062,2066                  & $12.69\pm 0.05$\phantom{$^{\rm a}$} & \phantom{uuu..}''      \\
Fe\,{\sc ii} & 1096,1125,1144 $^{\rm a}$       & $14.20\pm 0.02$\phantom{$^{\rm a}$} & \phantom{uuu..}''      \\
Ni\,{\sc ii} & 1741,1751                       & $13.07\pm 0.09$\phantom{$^{\rm a}$} & \phantom{uuu..}''      \\
Zn\,{\sc ii} & 2026,2062                       & $11.81\pm 0.12$\phantom{$^{\rm a}$} & \phantom{uuu..}''      \\
\hline
\multicolumn{4}{l}{$z_{\rm abs}=3.02485$}\\
C\,{\sc i}                   & 1656 $^{\rm f}$ & $11.73\pm 0.26$\phantom{$^{\rm a}$} & \phantom{uuu.}...      \\
C\,{\sc i}\,$^\star$         & 1656            & $<11.75$ $^{\rm e}$                 & \phantom{uuu.}...      \\
C\,{\sc i}\,$^{\star\star }$ & 1657            & $<11.50$ $^{\rm e}$                 & \phantom{uuu.}...      \\
C\,{\sc ii}\,$^\star$        & 1037            & $13.55\pm 0.23$\phantom{$^{\rm a}$} & \phantom{0}$4.9\pm 0.3$\\
N\,{\sc i}   & 1134.4,1134.9,1200.2            & $14.47\pm 0.03$\phantom{$^{\rm a}$} & \phantom{uuu..}''      \\
O\,{\sc i}   & 950,974                         & $16.18\pm 0.18$ $^{\rm b}$          & \phantom{uuu..}''      \\
Si\,{\sc ii} & 1020,1808                       & $14.48\pm 0.04$\phantom{$^{\rm a}$} & \phantom{uuu..}''      \\
P\,{\sc ii}  & 963                             & $12.48\pm 0.09$ $^{\rm c}$          & \phantom{uuu..}''      \\
S\,{\sc ii}  & 1259 $^{\rm d}$                 & $14.36\pm 0.04$\phantom{$^{\rm a}$} & \phantom{uuu..}''      \\
Ar\,{\sc i}  & 1048                            & $13.44\pm 0.05$\phantom{$^{\rm a}$} & \phantom{uuu..}''      \\
Ti\,{\sc ii} & 1910.6,1910.9                   & $<11.86$ $^{\rm e}$                 & \phantom{uuu..}''      \\
Cr\,{\sc ii} & 2056,2062,2066                  & $12.29\pm 0.11$\phantom{$^{\rm a}$} & \phantom{uuu..}''      \\
Fe\,{\sc ii} & 1096,1125,1144 $^{\rm a}$       & $13.81\pm 0.05$\phantom{$^{\rm a}$} & \phantom{uuu..}''      \\
Ni\,{\sc ii} & 1741,1751                       & $12.65\pm 0.19$\phantom{$^{\rm a}$} & \phantom{uuu..}''      \\
Zn\,{\sc ii} & 2026,2062                       & $12.02\pm 0.04$\phantom{$^{\rm a}$} & \phantom{uuu..}''      \\
\hline
\multicolumn{4}{l}{$z_{\rm abs}=3.02501$}\\
N\,{\sc i}   & 1134.4,1134.9,1200.2            & $13.44\pm 0.04$\phantom{$^{\rm a}$} & \phantom{0}$6.1\pm 0.5$\\
S\,{\sc ii}  & 1259 $^{\rm d}$                 & $13.51\pm 0.12$\phantom{$^{\rm a}$} & \phantom{uuu..}''      \\
Ar\,{\sc i}  & 1048                            & $12.71\pm 0.06$\phantom{$^{\rm a}$} & \phantom{uuu..}''      \\
Cr\,{\sc ii} & 2056,2062,2066                  & $12.25\pm 0.09$\phantom{$^{\rm a}$} & \phantom{uuu..}''      \\
Fe\,{\sc ii} & 1096,1125,1144 $^{\rm a}$       & $13.69\pm 0.03$\phantom{$^{\rm a}$} & \phantom{uuu..}''      \\
Ni\,{\sc ii} & 1741,1751                       & $12.56\pm 0.17$\phantom{$^{\rm a}$} & \phantom{uuu..}''      \\
\hline
\end{tabular}
\label{tabmet2}
\flushleft
$^{\rm a}$ Also 1608, 1611.\\
$^{\rm b}$ An additional error of 20 percent was incorporated to reflect
the uncertainty on the oscillator strength of the O\,{\sc i}\,$\lambda$974
line (see Morton 1991).\\
$^{\rm c}$ The quoted error takes into account the main uncertainty which
is related to the continuum placement because the single line used is
located in a crowded Lyman-$\alpha$ forest.\\
$^{\rm d}$ Line observed by Prochaska \& Wolfe (1999).\\
$^{\rm e}$ $5\sigma$ upper limit.\\
$^{\rm f}$ A $3\sigma$ absorption feature is detected at
$z_{\rm abs}=3.02490$ ($\sim 3$ km s$^{-1}$ redward of $z_{\rm abs}=3.02485$)
while the H$_2$ lines are consistently fitted with a single component
at $z_{\rm abs}=3.02489$ (see Table~\ref{tabmol2}).
\end{table}

The detection of molecular hydrogen toward this quasar has been
reported recently by Levshakov et al. (2002) who used UVES commissioning data.
As part of our systematic survey, we have obtained significantly better UVES
data for this line of sight and we reanalyse here the DLA system at
$z_{\rm abs}=3.025$ in the same footing as the rest of our sample.

We display in Fig.~\ref{figmetals2} the velocity profiles of the
most important lines from low ions and neutral species that are present in our
data. Two well-separated components are observed for a number of ions,
namely N\,{\sc i}, O\,{\sc i}, Si\,{\sc ii}, P\,{\sc ii}, S\,{\sc ii} and
Ar\,{\sc i}. One of these two components, at $z_{\rm abs}=3.02485$, has
detected H$_2$ lines.

\begin{figure}
\centerline{\hbox{
\psfig{figure=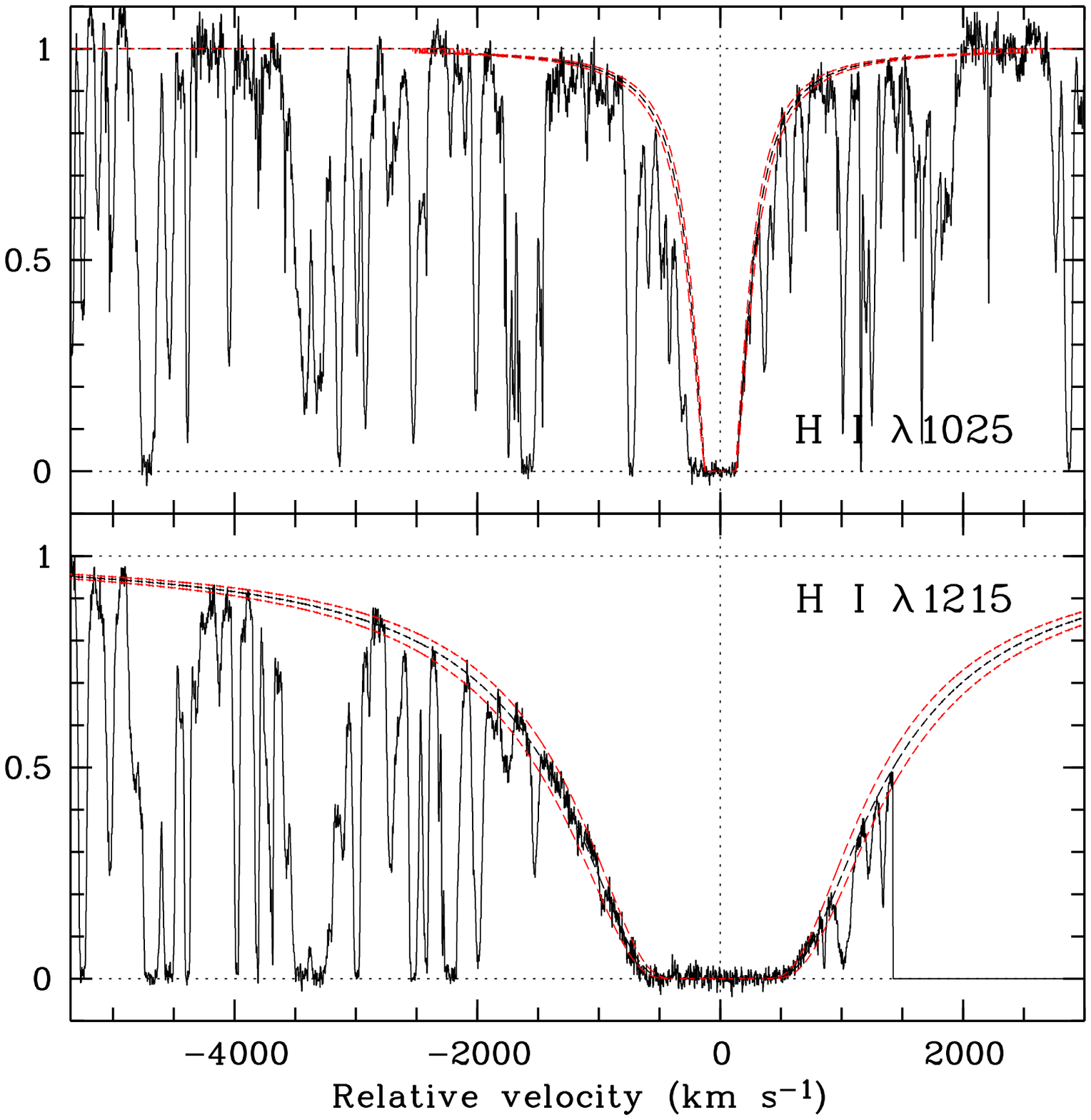,width=8.4cm,clip=,bbllx=53.pt,bblly=41.pt,bburx=560.pt,bbury=542.pt,angle=0.}}}
\caption[]{Portions of the normalized UVES spectrum showing the
damped Ly$\alpha$ and Ly$\beta$ lines of the DLA system at
$z_{\rm abs}=3.025$ toward Q\,0347$-$383. The best-fitted Voigt
profiles superimposed on the data correspond
to $N($H\,{\sc i}$)=(3.6\pm 0.4)\times 10^{20}$ cm$^{-2}$.}
\label{figlya2}
\end{figure}

Fitted altogether (see below), the detected H$_2$ lines are found to
be consistently located at a redshift of 3.02489 and show a systematic
offset of $\sim 3$ km s$^{-1}$ relative to the deepest part of
the low-ion profiles. We simultaneously fitted the lines from
different rotational levels using the same broadening parameter and redshift,
and measured accurate column densities for the $J=0$ to 4 levels
from sub-samples of 4 to 7 unblended lines per $J$ level. The results
are given in Table~\ref{tabmol2} and the best-fitting model is shown
in Fig.~\ref{figmole2}. Note in particular that the measurement of the column
density in $J=0$ is based on the detection at more than $5\sigma$ of four
transition lines: W\,1-0\,R(0), L\,7-0\,R(0), L\,3-0\,R(0) and L\,2-0\,R(0). A
$3\sigma$ absorption feature, possibly corresponding
to C\,{\sc i}\,$\lambda$1656, is also detected at a redshift consistent
with that of the H$_2$ lines. If this is true then
$\log N($C\,{\sc i}$)=11.73\pm 0.26$ (see Table~\ref{tabmet2}). The second
low-ion line component, at $z_{\rm abs}=3.02463$, has no detected H$_2$ down
to, respectively, $\log N($H$_2)<13.0$ and $<13.4$ ($5\sigma$ limits)
for, respectively, $J=0$ and 1. We measured in this DLA system a total
neutral hydrogen column density $\log N($H\,{\sc i}$)=20.56\pm 0.05$ (see
Fig.~\ref{figlya2}) and a total molecular hydrogen column density
$\log N($H$_2)=14.55\pm 0.09$ (see Table~\ref{tabmol2}). This leads to an
overall molecular fraction, $\log f=-5.71\pm 0.10$, in agreement with that
derived by Levshakov et al. (2002).

We measured ionic column densities in the two main components of the
system and in two weaker ones only for species with unsaturated and/or mildly
saturated lines. The results are given in Table~\ref{tabmet2}. The
broadening parameters were mainly constrained by Fe\,{\sc ii} and, to
some extent also, N\,{\sc i} lines covering a range of oscillator strengths.
It can be seen in Fig.~\ref{figmetals2} that in our data the Zn\,{\sc ii}
doublet lines are firmly detected in between sky lines in the strongest
component of the system ($z_{\rm abs}=3.02485$) and are also likely present in
the component at $z_{\rm abs}=3.02463$. In addition, the Cr\,{\sc ii} triplet
is nicely detected in each of the four considered components
(see Fig.~\ref{figmetals2} and Table~\ref{tabmet2}).

In Fig.~\ref{figmetals2}, we show evidence that the
Si\,{\sc ii}\,$\lambda$1808 line is badly blended with a telluric absorption
feature which we have carefully removed via modelling. The best-fitting of
the S\,{\sc ii}\,$\lambda$1259 line from the Keck-HIRES spectrum of
Prochaska \& Wolfe (1999; private communication) and of
both Si\,{\sc ii}\,$\lambda\lambda$1020,1808 lines in our VLT-UVES
spectrum implies that the $\alpha$- over iron-peak elemental ratio
derived from the two main components of the system is close to Solar and
may even be slightly under-Solar:
[S,~Si/Zn$]=-0.11\pm 0.09,-0.35\pm 0.09$. This is line with the findings of
studies of large DLA
samples (Prochaska \& Wolfe 2002; Ledoux, Bergeron \& Petitjean 2002a). In
addition, we report the first detection of the O\,{\sc i}\,$\lambda$974
line in a DLA system (at the $4\sigma$ significance level;
see Fig.~\ref{figmetals2}). The vacuum wavelength and the oscillator
strength of this line are poorly-known quantities however (see Morton 1991).
In order to fit the line, we used $\lambda _{\rm vac}=974.075$ \AA , and we
included at the end an extra 20 percent uncertainty on the measured O\,{\sc i}
column density to reflect the uncertainty on the oscillator strength.
We obtain [O/Zn$]=0.00\pm 0.20$. Being Solar, this ratio is in agreement
with the ones previously derived for S and Si. This is at variance with
the claim by Levshakov et al. (2002) that in this DLA system $\alpha$-elements
are strongly overabundant compared to zinc ([$\alpha$/Zn$]=0.6\pm 0.1$). The
latter authors have underestimated both the contamination of metal lines
by telluric absorption features at $\lambda _{\rm obs}>7000$ \AA\ and also the
saturation of O\,{\sc i} lines in the Lyman-$\alpha$ forest.

\begin{figure*}
\flushleft{\vbox{
\psfig{figure=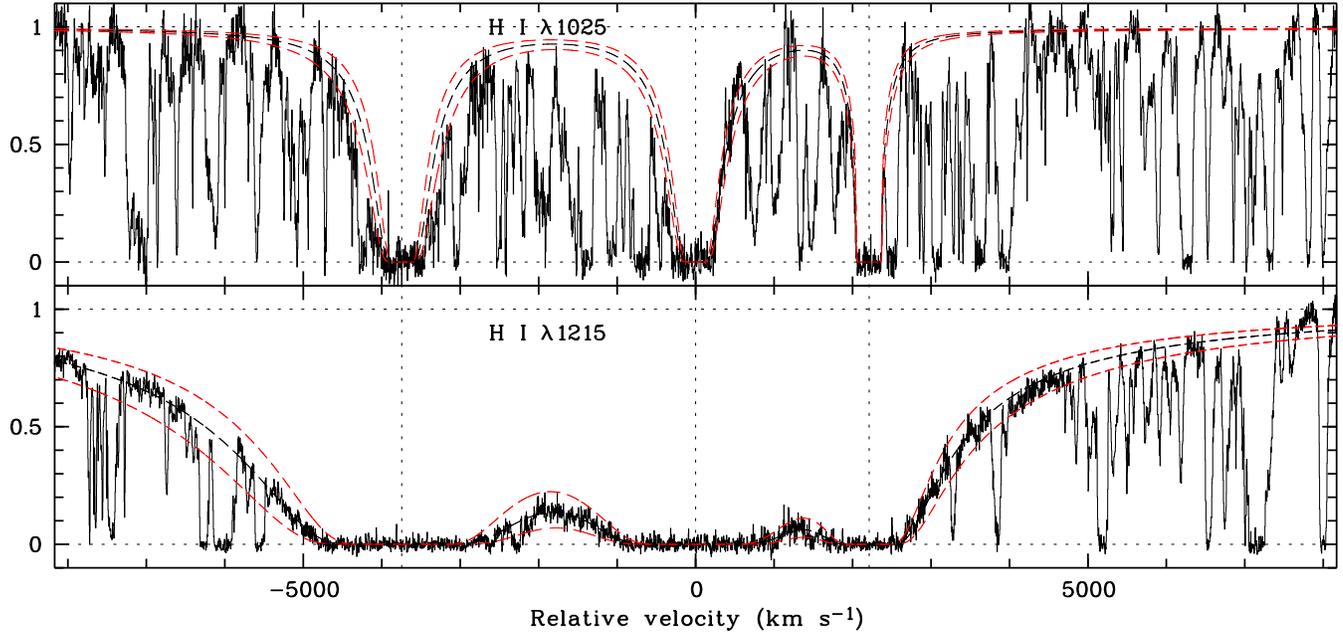,width=17.7cm,clip=,bbllx=204.pt,bblly=60.pt,bburx=546.pt,bbury=769.pt,angle=270.}}}
\caption[]{Portions of the normalized UVES spectrum showing the
damped Ly$\alpha$ and Ly$\beta$ lines of the DLA/sub-DLA systems at
$z_{\rm abs}=2.550$, 2.595 and 2.621 (see vertical lines) toward
Q\,0405$-$443. The best-fitted Voigt profiles superimposed on the
data correspond
to $N($H\,{\sc i}$)=(10\pm 4,7.9\pm 2.1,1.8\pm 0.5)\times 10^{20}$ cm$^{-2}$
for each of the systems respectively.}
\label{figlya4}
\end{figure*}

\begin{figure*}
\flushleft{\hbox{
\psfig{figure=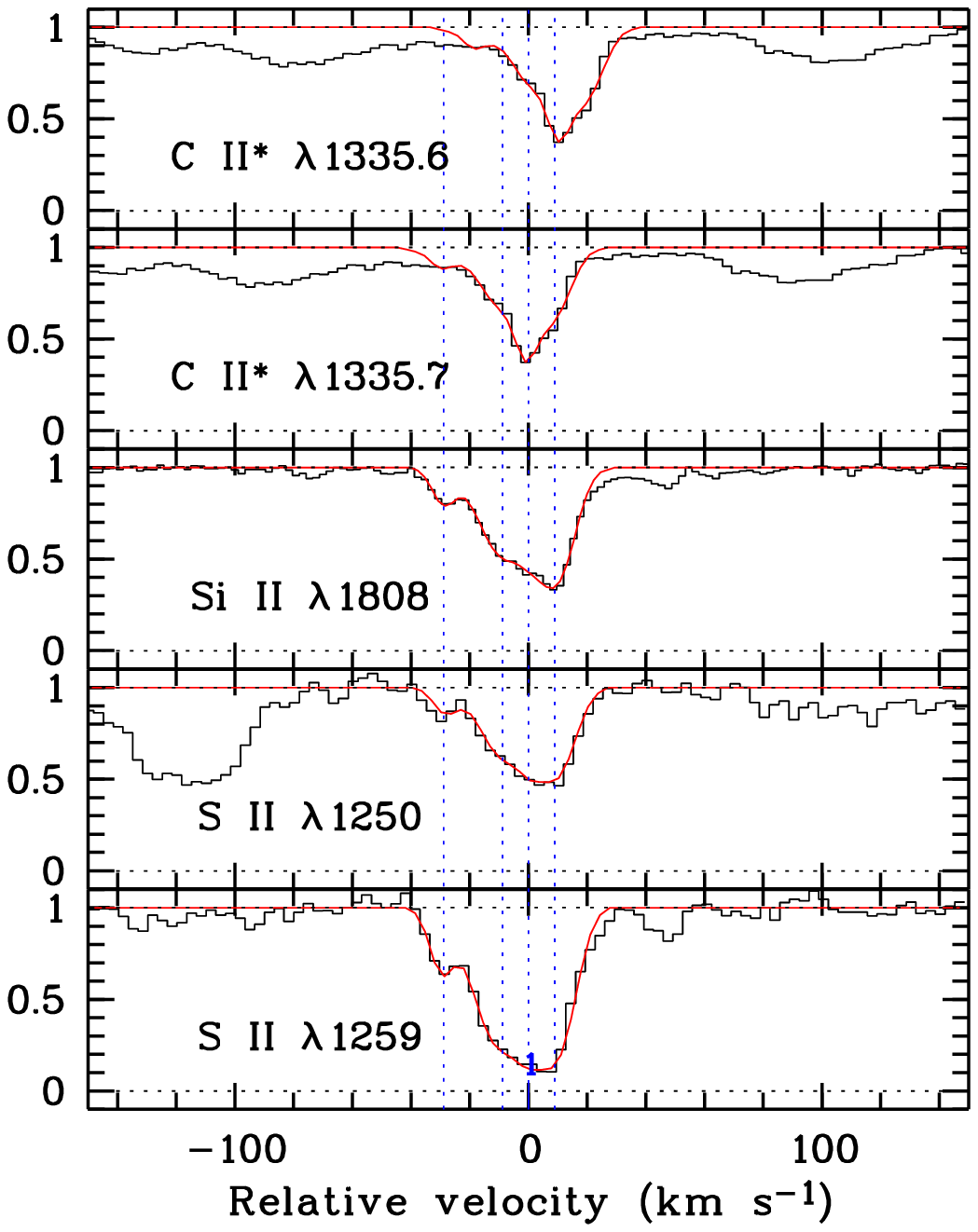,width=8.65cm,clip=,bbllx=57.pt,bblly=388.pt,bburx=360.pt,bbury=768.pt,angle=0.}\hspace{+0.3cm}\psfig{figure=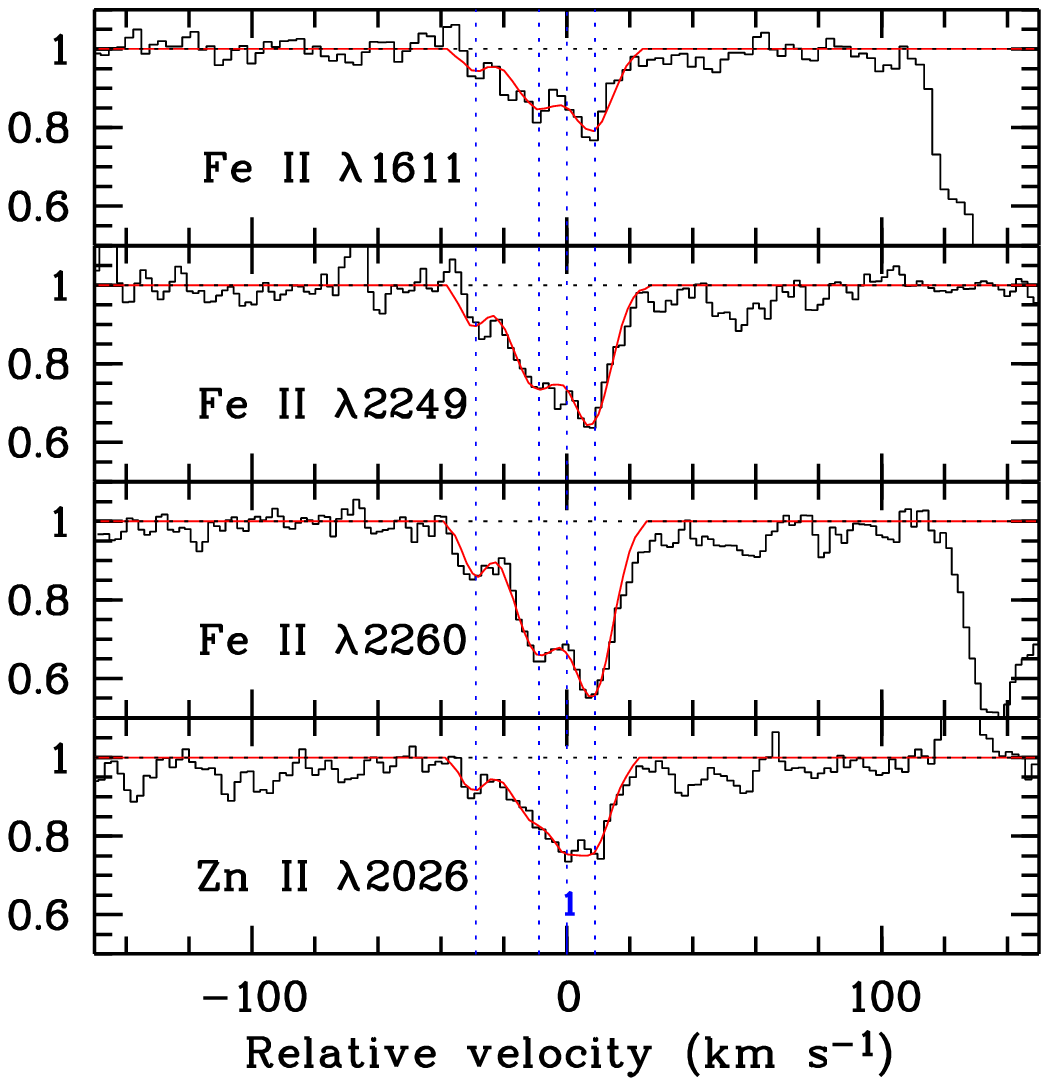,width=8.65cm,clip=,bbllx=57.pt,bblly=456.pt,bburx=360.pt,bbury=837.pt,angle=0.}}}
\caption[]{Velocity profiles of some of the low-ion lines in the DLA system at
$z_{\rm abs}=2.595$ toward Q\,0405$-$443. Our best-fitting model
is superimposed on the spectra with vertical lines marking the location of
individual components. The single component where H$_2$ is detected is labelled
with number 1. This component is clearly seen in the
C\,{\sc ii}\,$^\star\lambda$1335.7 transition line and, to some extent also,
in the S\,{\sc ii} and Zn\,{\sc ii} profiles.}
\label{figmetals4}
\end{figure*}

We compute metallicities [Zn/H$]=-0.98\pm 0.09$ and [S/H$]=-1.09\pm 0.06$,
and a mean depletion factor [Zn/Fe$]=0.74\pm 0.09$, using the
integrated column densities from the two main components of the system at
$z_{\rm abs}=3.02463$ and 3.02485. Interestingly, the depletion,
supposedly onto dust grains, and maybe also the metallicity, is larger in
the component where H$_2$ is
detected ([Zn,~S/Fe$]=1.07\pm 0.06$,~$0.79\pm 0.06$ at $z_{\rm abs}=3.02485$)
relative to the other components ([Zn,~S/Fe$]=0.47\pm 0.12$,~$0.54\pm 0.04$
and [S/Fe$]=0.06\pm 0.12$ at, respectively, $z_{\rm abs}=3.02463$ and
3.02501). Fig.~\ref{figmetals2} indeed strikingly shows that the
absorption line profiles of little or non-depleted elements, such as Si, P and
S, are different from, e.g., the Cr and Fe profiles in their being about twice
as narrow.

\subsection{Q\,0405$-$443, $z_{\rm abs}=2.595$}\label{q0405}

Lopez et al. (2001) reported the discovery of three DLA candidates in
a low-resolution spectrum of Q\,0405$-$443. We confirm here from UVES
spectroscopy the damped nature of these absorption systems at redshifts
$z_{\rm abs}=2.550$, 2.595 and 2.621. Because the continuum of this quasar
is especially well defined in our high-resolution spectra, we are able to
derive accurate total neutral hydrogen column densities for each absorber.
For this, we fitted the Lyman series from Ly$\alpha$ to Ly6 and
obtain $\log N($H\,{\sc i}$)=21.00\pm 0.15$, $20.90\pm 0.10$ and
$20.25\pm 0.10$ at, respectively, the above redshifts. Fig.~\ref{figlya4}
shows the best-fitting to the data which is in practice mainly constrained by
the combination of Ly$\alpha$ and Ly$\beta$. A complete analysis of the three
absorption systems will be published in Lopez et al. (in prep.) who have
independently acquired UVES data of a similar quality for the purpose of
studying elemental abundances. We concentrate here on only some of the metal
lines observed in the DLA system at $z_{\rm abs}=2.595$
(see Fig.~\ref{figmetals4}) as in this particular absorber we found H$_2$.

\begin{figure*}
\flushleft{\vbox{
\psfig{figure=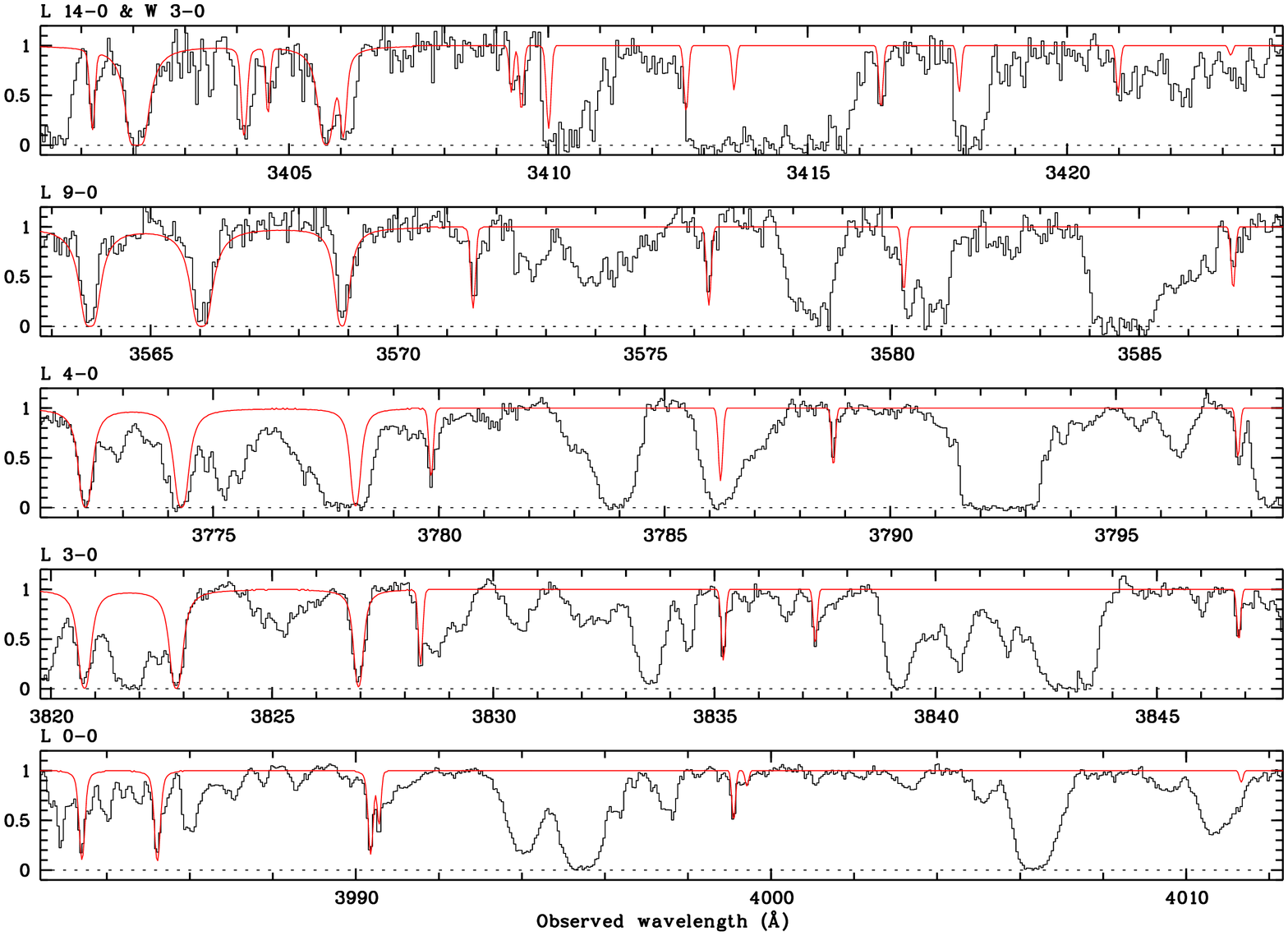,width=17.7cm,clip=,bbllx=41.pt,bblly=63.pt,bburx=555.pt,bbury=778.pt,angle=270.}}}
\caption[]{Voigt-profile fitting to the transition lines from the $J=0$, 1,
2 and 3 rotational levels of the vibrational ground-state Lyman and
Werner bands of H$_2$ at $z_{\rm abs}=2.59471$ toward Q\,0405$-$443. Only a
few of the observed bands are displayed. H$_2$ is detected on this line
of sight in a single gas cloud.}
\label{figmole4}
\end{figure*}

Molecular hydrogen at $z_{\rm abs}=2.595$ is detected in the $J=0$, 1, 2 and 3
rotational levels. Interestingly, quite a large number of H$_2$ lines from
the $L=0$ to 14 Lyman-bands is observed at signal-to-noise ratios of the
order of 10 to 30 from $\lambda _{\rm obs}=3500$ to 4000 \AA . This leads to a
precise determination of the molecular hydrogen column densities in $J=0$ and
1, and to some extent also, in $J=2$ and 3 (see below however).
Unsaturated and/or only moderately strong lines are observed for
all $J\le 3$ and damping wings are present in Werner-band lines from $J=0$ and
1. We simultaneously fitted all unblended lines from different rotational
levels, using the same broadening parameter and redshift, with a
single component at $z_{\rm abs}=2.59471$. The column density in each $J$
level was derived from several trials of Voigt-profile fitting to take
into account the range of possible $b$ values. Errors in the column
densities therefore correspond to a range of column densities and not to
the rms error from fitting the Voigt profiles. The results are given
in Table~\ref{tabmol4} and the best-fitting model is shown
in Fig.~\ref{figmole4}. The broadening parameter is found to be $b=1.5$
km s$^{-1}$. We note however that the relative strengths of the $J\ge 2$
lines require $b\ga 1.5$ km s$^{-1}$ whereas the $J\le 1$ lines are consistent
with $b\la 1.5$ km s$^{-1}$. The broadening parameter could thus be
smaller for lines from lower $J$ and larger for lines from higher $J$, which
may affects the determination of the column densities only in $J\ge 2$.

\begin{table}
\caption {Voigt-profile fitting results for different rotational levels of the
vibrational ground-state Lyman and Werner bands of H$_2$ toward Q\,0405$-$443}
\begin{tabular}{llll}
\hline
\hline
$z_{\rm abs}$ & Level & $\log N\pm\sigma _{\log N}$ & $b\pm\sigma _b$\\
              &       & (H$_2$)                     & (km s$^{-1}$)  \\
\hline
2.59471 & $J=0$ & $17.73^{+0.15}_{-0.05}$ & $1.5\pm 0.5$    \\ %old+/-0.22
        & $J=1$ & $17.95^{+0.20}_{-0.05}$ & \phantom{uu..}''\\ %old+/-0.37
        & $J=2$ & $15.71^{+0.90}_{-0.49}$ & \phantom{uu..}''\\ %old+/-0.36
        & $J=3$ & $14.70^{+0.70}_{-0.22}$ & \phantom{uu..}''\\ %old+/-0.19
        & $J=4$ & $<13.87$ $^{\rm a}$     & \phantom{uu..}''\\
        & $J=5$ & $<13.50$ $^{\rm a}$     & \phantom{uu..}''\\
\hline
\end{tabular}
\label{tabmol4}
\flushleft Note: errors in the column densities correspond to a range
of column densities; they are not the rms errors from fitting the
Voigt profiles (see text).\\
$^{\rm a}$ $5\sigma$ upper limit.
\end{table}

The total molecular hydrogen column density measured in this system,
$\log N($H$_2)=18.16^{+0.21}_{-0.06}$, is one of the largest ever seen in DLA
systems (see Sect.~\ref{sam}). However, the corresponding molecular
fraction, $\log f=-2.44^{+0.23}_{-0.12}$, is relatively small due to the
fairly large total neutral hydrogen column density. In spite of the high
signal-to-noise ratio in our data at the location of the
C\,{\sc i}\,$\lambda$1656 transition line, neutral carbon is not
clearly detected. A $\ga 5\sigma$ absorption feature, whose origin is
uncertain, is present at $z_{\rm abs}=2.59485$, corresponding
to $\log N($C\,{\sc i}$)\le 11.95$.

We measured ionic column densities in the three main components of the
system and in a weaker one. The results are given in Table~\ref{tabmet4}. The
component where H$_2$ is detected is not seen at all in Fe\,{\sc ii}, but is
prominent in the C\,{\sc ii}\,$^\star$ profile and detected in Si\,{\sc ii},
S\,{\sc ii} and Zn\,{\sc ii} (see Fig.~\ref{figmetals4}). A careful
modelling of the central blend leads to an accurate determination of the
column densities of the latter ions, and to an upper limit on the column
density of Fe\,{\sc ii} in the component at $z_{\rm abs}=2.59474$ which
is likely associated with the H$_2$ absorption lines (see Table~\ref{tabmet4}).
We compute overall metallicities [Zn/H$]=-1.02\pm 0.12$ and
[S/H$]=-1.07\pm 0.10$, and a mean depletion factor [Zn/Fe$]=0.31\pm 0.08$. As
previously in the case of the DLA system toward Q\,0347$-$383, the
depletion is found to be larger in the component where H$_2$ is detected
([Zn/Fe$]>0.70$ at $z_{\rm abs}=2.59474$) relative to the other components
([Zn/Fe$]=0.38\pm 0.11$, $0.25\pm 0.06$ and $0.28\pm 0.05$ at, respectively,
$z_{\rm abs}=2.59440$, 2.59464 and 2.59485).

\subsection{Q\,1444$+$014, $z_{\rm abs}=2.087$}\label{q1444}

\begin{figure}
\flushleft{\vbox{
\psfig{figure=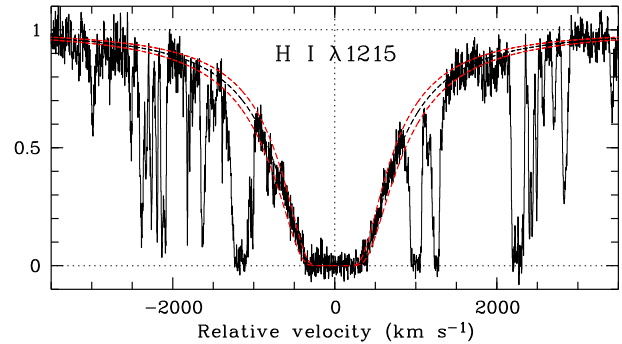,width=8.4cm,clip=,bbllx=53.pt,bblly=268.pt,bburx=560.pt,bbury=542.pt,angle=0.}}}
\caption[]{Portion of the normalized UVES spectrum showing the
damped Ly$\alpha$ line of the sub-DLA system at $z_{\rm abs}=2.087$ toward
Q\,1444$+$014. The best-fitted Voigt profile superimposed on the
data corresponds to $N($H\,{\sc i}$)=(1.2\pm 0.2)\times 10^{20}$ cm$^{-2}$.}
\label{figlya3}
\end{figure}

\begin{figure*}
\flushleft{\hbox{
\psfig{figure=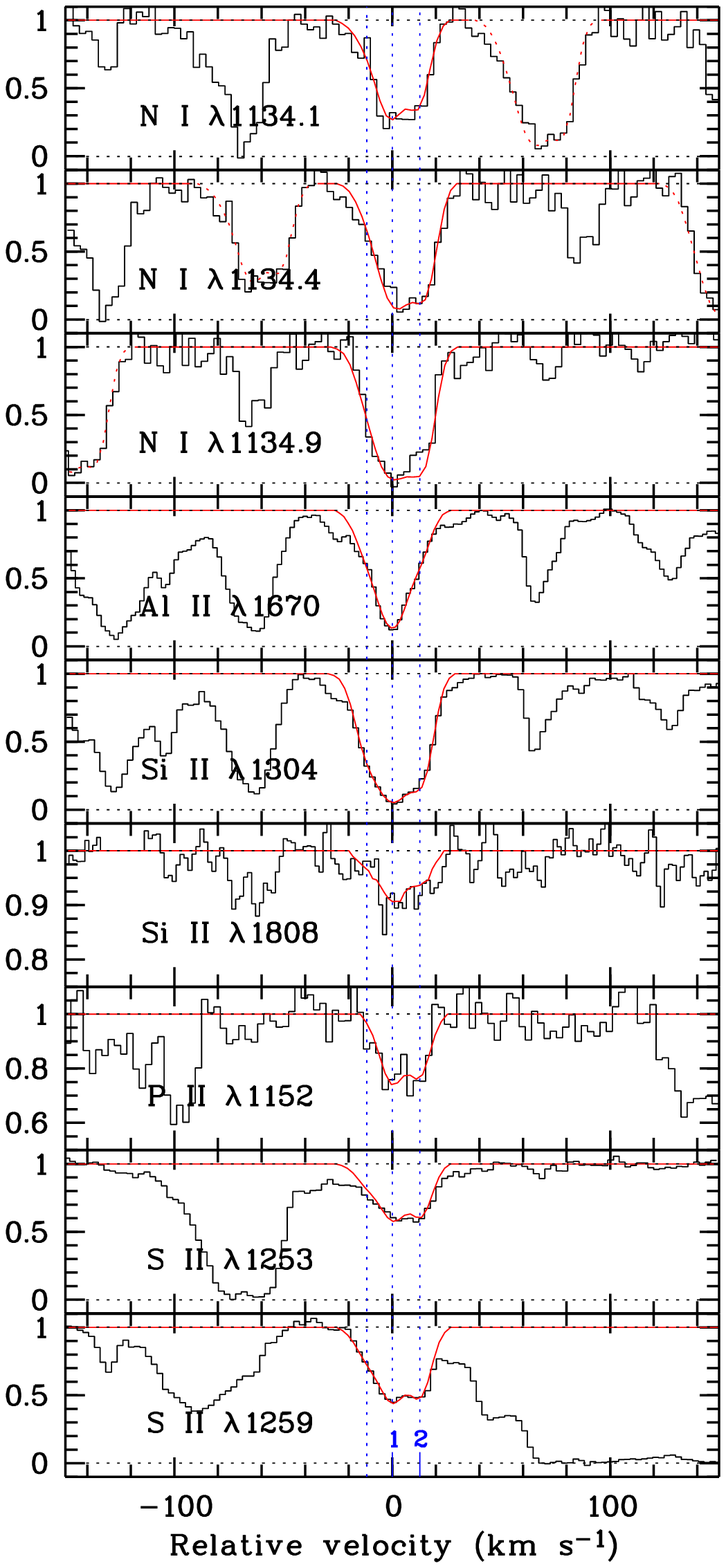,width=8.65cm,clip=,bbllx=57.pt,bblly=116.pt,bburx=360.pt,bbury=769.pt,angle=0.}\hspace{+0.3cm}\psfig{figure=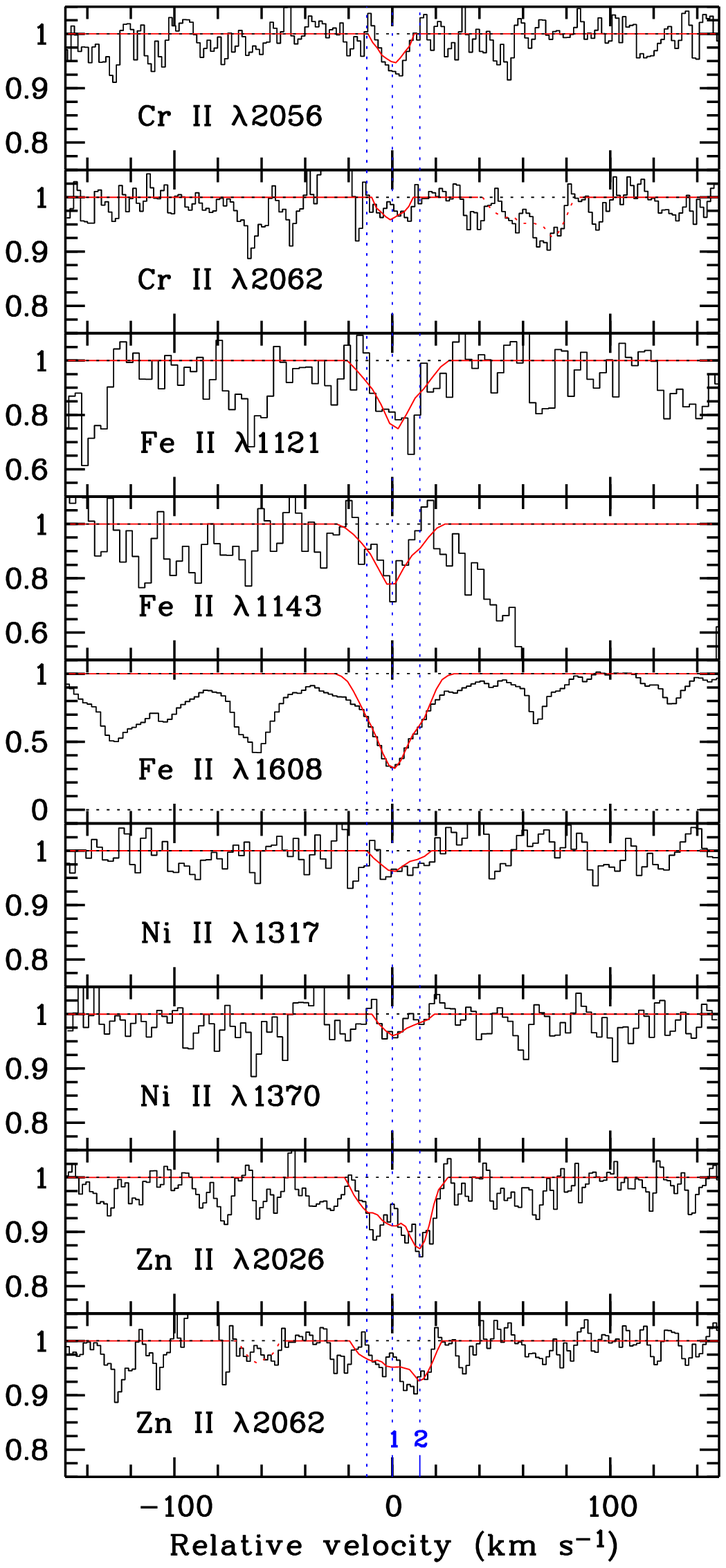,width=8.65cm,clip=,bbllx=57.pt,bblly=116.pt,bburx=360.pt,bbury=769.pt,angle=0.}}}
\caption[]{Velocity profiles of lines from low ions and neutral species in the
sub-DLA system at $z_{\rm abs}=2.087$ toward Q\,1444$+$014.
Our best-fitting model of the central part of the profiles is superimposed on
the spectra with vertical lines marking the location of individual
components. The dotted parts in some of the synthetic profiles correspond
to other transitions than the ones indicated. The two components
where H$_2$ is detected are labelled with number 1 and number 2 respectively.}
\label{figmetals3}
\end{figure*}

\begin{figure*}
\flushleft{\hbox{
\psfig{figure=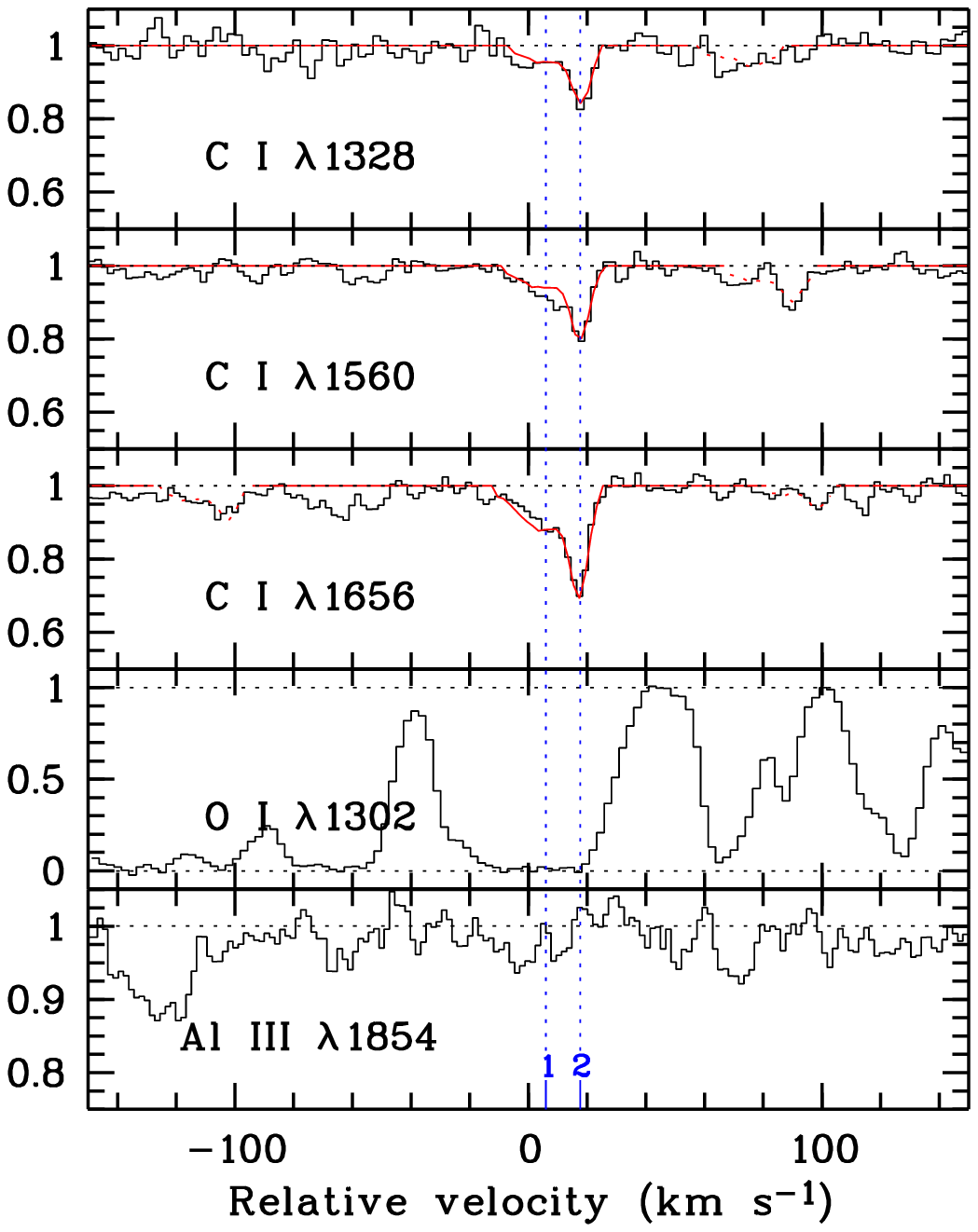,width=8.65cm,clip=,bbllx=57.pt,bblly=387.pt,bburx=360.pt,bbury=769.pt,angle=0.}\hspace{+0.3cm}\psfig{figure=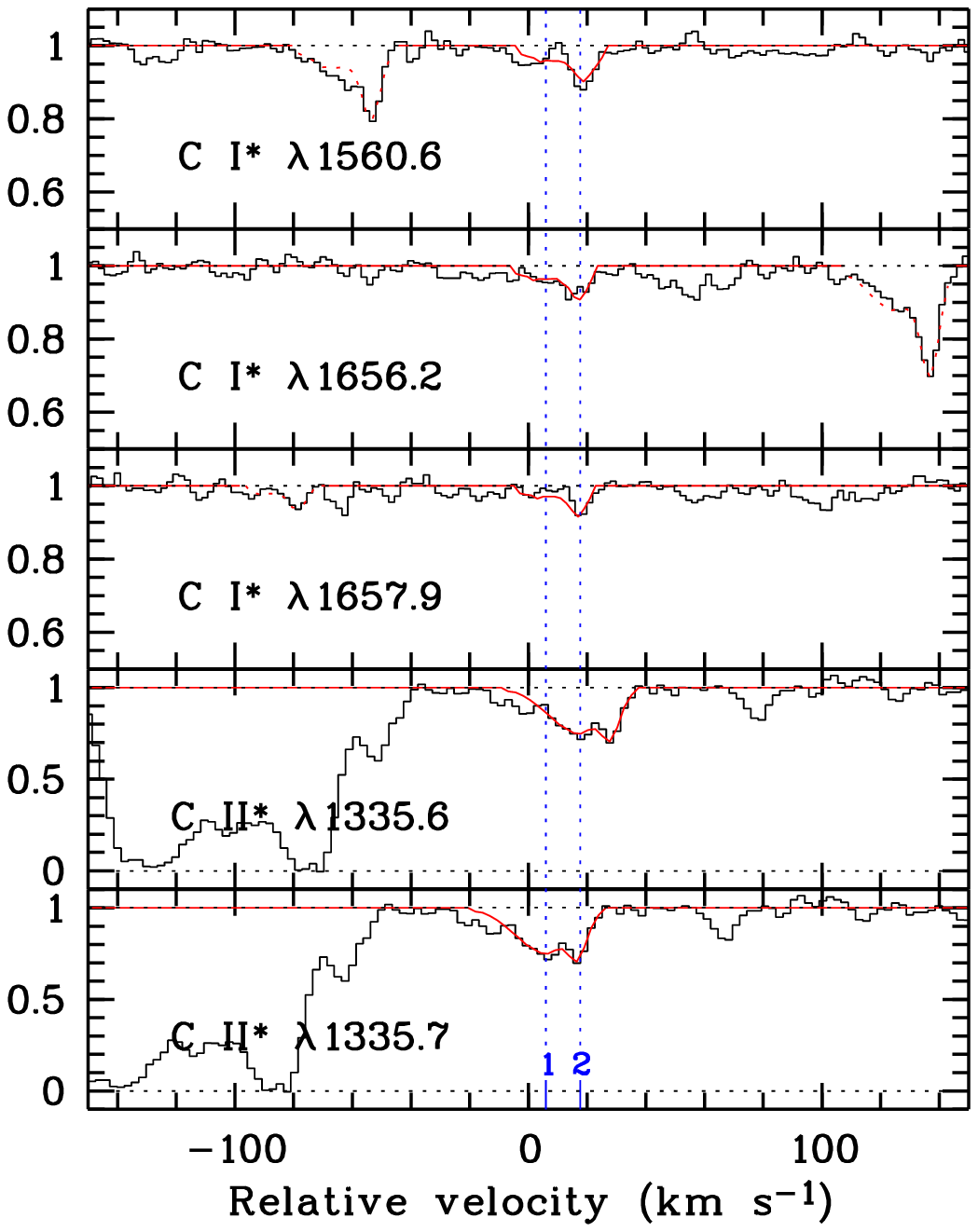,width=8.65cm,clip=,bbllx=57.pt,bblly=387.pt,bburx=360.pt,bbury=769.pt,angle=0.}}}
\caption[]{Absorption lines from the ground-state of C\,{\sc i} (upper part of
left panel) and fine-structure levels of C\,{\sc i} and C\,{\sc ii}
(right panels) at $z_{\rm abs}=2.087$ toward Q\,1444$+$014. The two fitted
components, where H$_2$ is detected, are indicated as in Fig.~\ref{figmetals3}
by numbers in addition to vertical lines. The dotted parts in some of the
synthetic profiles correspond to other transitions than the ones indicated.
The origin of the velocity scale is the redshift of the main component as
measured from the other metal lines ($z_{\rm abs}=2.08679$). The redward shift
of the C\,{\sc i} components is apparent. The velocity profile
of O\,{\sc i}\,$\lambda$1302 is shown for comparison. Note
that Al\,{\sc iii}\,$\lambda$1854 is barely detected (bottom part of left
panel) in this sub-DLA system.}
\label{figcarbon3}
\end{figure*}

Voigt-profile fitting to the Lyman-$\alpha$ line of this absorption system
leads to a total neutral hydrogen column density
$\log N($H\,{\sc i}$)=20.07\pm 0.07$ (see Fig.~\ref{figlya3}). Although this
system would thus not qualify as a DLA system following the conventional
definition, strong associated H$_2$ lines are detected.
Moreover, Al\,{\sc iii}, Si\,{\sc iv} and C\,{\sc iv} lines are
barely detected at $z_{\rm abs}\approx 2.087$ which shows that the amount of
highly ionized gas is small in the system. The overall metal line profiles are
unusual in displaying a series of well-separated sub-systems spread over
$\sim 400$ km s$^{-1}$. However, the N\,{\sc i}, P\,{\sc ii}
and Zn\,{\sc ii} profiles suggest that most of the neutral hydrogen is
concentrated within $\sim 40$ km s$^{-1}$ of the central clump
(see Fig.~\ref{figmetals3}). This is confirmed by the detection of
several C\,{\sc i} lines from different excitation states in two components at
$z_{\rm abs}=2.08685$ and 2.08697 (velocity separation 12 km s$^{-1}$; see
Fig.~\ref{figcarbon3}). The component at $z_{\rm abs}=2.08697$ is narrow
and unresolved; the measured $b$-value is 1.1 km s$^{-1}$ (see
Table~\ref{tabmet3}).

\begin{table}
\caption {Ionic column densities in individual components of the DLA system
at $z_{\rm abs}=2.595$ toward Q\,0405$-$443}
\begin{tabular}{llll}
\hline
\hline
Ion & Transition & $\log N\pm\sigma _{\log N}$ & $b\pm\sigma _b$\\
    & lines used &                             & (km s$^{-1}$)  \\
\hline
\multicolumn{4}{l}{$z_{\rm abs}=2.59440$}           \\
C\,{\sc ii}\,$^\star$        & 1335.6,1335.7             & $\le 12.41\pm 0.14$\phantom{$^{\rm a}$}  & $3.7\pm 0.2$\phantom{:$^{\rm b}$}\\
Si\,{\sc ii}                 & 1808                      & $14.35\pm 0.05$\phantom{$^{\rm a}$}      & \phantom{uu..}''                 \\
S\,{\sc ii}                  & 1250,1259                 & $13.91\pm 0.04$\phantom{$^{\rm a}$}      & \phantom{uu..}''                 \\
Fe\,{\sc ii}                 & 2249,2260,2374 $^{\rm a}$ & $13.98\pm 0.02$\phantom{$^{\rm a}$}      & \phantom{uu..}''                 \\
Zn\,{\sc ii}                 & 2026                      & $11.50\pm 0.11$\phantom{$^{\rm a}$}      & \phantom{uu..}''                 \\
\hline
\multicolumn{4}{l}{$z_{\rm abs}=2.59464$}           \\
C\,{\sc ii}\,$^\star$        & 1335.6,1335.7             & $13.09\pm 0.10$\phantom{$^{\rm a}$}      & $9.8\pm 0.2$\phantom{:$^{\rm b}$}\\
Si\,{\sc ii}                 & 1808                      & $15.11\pm 0.02$\phantom{$^{\rm a}$}      & \phantom{uu..}''                 \\
S\,{\sc ii}                  & 1250,1259                 & $14.72\pm 0.02$\phantom{$^{\rm a}$}      & \phantom{uu..}''                 \\
Fe\,{\sc ii}                 & 2249,2260,2374 $^{\rm a}$ & $14.72\pm 0.04$\phantom{$^{\rm a}$}      & \phantom{uu..}''                 \\
Zn\,{\sc ii}                 & 2026                      & $12.11\pm 0.04$\phantom{$^{\rm a}$}      & \phantom{uu..}''                 \\
\hline
\multicolumn{4}{l}{$z_{\rm abs}=2.59474$ $^{\rm b}$}\\
C\,{\sc i}                   & 1656                      & $<11.84$ $^{\rm c}$\phantom{$\pm 0.00$}  & \phantom{uu.}...                 \\
C\,{\sc i}\,$^\star$         & 1656                      & $<11.95$ $^{\rm c}$\phantom{$\pm 0.00$}  & \phantom{uu.}...                 \\
C\,{\sc i}\,$^{\star\star }$ & 1657                      & $<11.82$ $^{\rm c}$\phantom{$\pm 0.00$}  & \phantom{uu.}...                 \\
C\,{\sc ii}\,$^\star$        & 1335.6,1335.7             & $13.20\pm 0.20$\phantom{$^{\rm a}$}      & \phantom{uu}$2.9$:               \\
Si\,{\sc ii}                 & 1808                      & $14.16\pm 0.14$\phantom{$^{\rm a}$}      & \phantom{uu..}''                 \\
S\,{\sc ii}                  & 1250,1259                 & $14.07\pm 0.09$\phantom{$^{\rm a}$}      & \phantom{uu..}''                 \\
Fe\,{\sc ii}                 & 2249,2260,2374 $^{\rm a}$ & $<13.68$ $^{\rm d}$\phantom{$\pm 0.00$}  & \phantom{uu..}''                 \\
Zn\,{\sc ii}                 & 2026                      & $11.52\pm 0.13$\phantom{$^{\rm a}$}      & \phantom{uu..}''                 \\
\hline
\multicolumn{4}{l}{$z_{\rm abs}=2.59485$}           \\
C\,{\sc i}                   & 1656                      & $\le 11.95\pm 0.34$:\phantom{$^{\rm a}$} & \phantom{uu.}...                 \\
C\,{\sc i}\,$^\star$         & 1656                      & $<11.95$ $^{\rm c}$\phantom{$\pm 0.00$}  & \phantom{uu.}...                 \\
C\,{\sc i}\,$^{\star\star }$ & 1657                      & $<11.82$ $^{\rm c}$\phantom{$\pm 0.00$}  & \phantom{uu.}...                 \\
C\,{\sc ii}\,$^\star$        & 1335.6,1335.7             & $13.25\pm 0.08$\phantom{$^{\rm a}$}      & $7.2\pm 0.3$\phantom{:$^{\rm b}$}\\
Si\,{\sc ii}                 & 1808                      & $15.18\pm 0.02$\phantom{$^{\rm a}$}      & \phantom{uu..}''                 \\
S\,{\sc ii}                  & 1250,1259                 & $14.73\pm 0.02$\phantom{$^{\rm a}$}      & \phantom{uu..}''                 \\
Fe\,{\sc ii}                 & 2249,2260,2374 $^{\rm a}$ & $14.75\pm 0.04$\phantom{$^{\rm a}$}      & \phantom{uu..}''                 \\
Zn\,{\sc ii}                 & 2026                      & $12.17\pm 0.03$\phantom{$^{\rm a}$}      & \phantom{uu..}''                 \\
\hline
\end{tabular}
\label{tabmet4}
\flushleft
$^{\rm a}$ Also 1608, 1611.\\
$^{\rm b}$ This component is likely associated with the H$_2$ absorption
lines observed at $z_{\rm abs}=2.59471$, 2.5 km s$^{-1}$ blueward
of $z_{\rm abs}=2.59474$.\\
$^{\rm c}$ $5\sigma$ upper limit.\\
$^{\rm d}$ Blend.
\end{table}

\begin{table}
\caption {Ionic column densities in individual components of the sub-DLA
system at $z_{\rm abs}=2.087$ toward Q\,1444$+$014}
\begin{tabular}{llll}
\hline
\hline
Ion & Transition & $\log N\pm\sigma _{\log N}$ & $b\pm\sigma _b$\\
    & lines used &                             & (km s$^{-1}$)  \\
\hline
\multicolumn{4}{l}{$z_{\rm abs}=2.08667$}\\
N\,{\sc i}   & 1134.1,1134.4,1134.9                   & $13.70\pm 0.10$\phantom{$^{\rm a}$} & \phantom{0}$6.1\pm 0.9$\phantom{$^{\rm b}$}\\
Al\,{\sc ii} & 1670                                   & $11.77\pm 0.07$\phantom{$^{\rm a}$} & \phantom{uuu..}''                          \\
Si\,{\sc ii} & 1304,1526,1808                         & $13.55\pm 0.08$\phantom{$^{\rm a}$} & \phantom{uuu..}''                          \\
S\,{\sc ii}  & 1253,1259                              & $13.79\pm 0.11$\phantom{$^{\rm a}$} & \phantom{uuu..}''                          \\
Fe\,{\sc ii} & 1121,1143,1608                         & $13.18\pm 0.09$\phantom{$^{\rm a}$} & \phantom{uuu..}''                          \\
Zn\,{\sc ii} & 2026,2062                              & $11.43\pm 0.18$\phantom{$^{\rm a}$} & \phantom{uuu..}''                          \\
\hline
\multicolumn{4}{l}{$z_{\rm abs}=2.08679$}\\
C\,{\sc i}            & 1328,1560,1656                & $12.67\pm 0.09$\phantom{$^{\rm a}$} & $11.6\pm 2.8$\phantom{$^{\rm b}$}          \\
C\,{\sc i}\,$^\star$  & 1656,1657.3,1657.9 $^{\rm a}$ & $12.52\pm 0.14$\phantom{$^{\rm a}$} & \phantom{uuu..}''                          \\
C\,{\sc ii}\,$^\star$ & 1335.6,1335.7                 & $13.12\pm 0.08$\phantom{$^{\rm a}$} & \phantom{uuu..}''                          \\
N\,{\sc i}   & 1134.1,1134.4,1134.9                   & $14.67\pm 0.04$\phantom{$^{\rm a}$} & \phantom{0}$6.9\pm 0.6$\phantom{$^{\rm b}$}\\
Mg\,{\sc ii} & 1239,1240                              & $<14.61$ $^{\rm b}$                 & \phantom{uuu..}''                          \\
Al\,{\sc ii} & 1670                                   & $12.56\pm 0.03$\phantom{$^{\rm a}$} & \phantom{uuu..}''                          \\
Si\,{\sc ii} & 1304,1526,1808                         & $14.15\pm 0.06$\phantom{$^{\rm a}$} & \phantom{uuu..}''                          \\
P\,{\sc ii}  & 1152                                   & $12.76\pm 0.06$\phantom{$^{\rm a}$} & \phantom{uuu..}''                          \\
S\,{\sc ii}  & 1253,1259                              & $14.32\pm 0.05$\phantom{$^{\rm a}$} & \phantom{uuu..}''                          \\
Ti\,{\sc ii} & 1910.6,1910.9                          & $<12.10$ $^{\rm b}$                 & \phantom{uuu..}''                          \\
Cr\,{\sc ii} & 2056,2062,2066                         & $12.14\pm 0.12$\phantom{$^{\rm a}$} & \phantom{uuu..}''                          \\
Fe\,{\sc ii} & 1121,1143,1608                         & $13.83\pm 0.04$\phantom{$^{\rm a}$} & \phantom{uuu..}''                          \\
Ni\,{\sc ii} & 1317,1370                              & $12.31\pm 0.21$\phantom{$^{\rm a}$} & \phantom{uuu..}''                          \\
Zn\,{\sc ii} & 2026,2062                              & $11.66\pm 0.12$\phantom{$^{\rm a}$} & \phantom{uuu..}''                          \\
\hline
\multicolumn{4}{l}{$z_{\rm abs}=2.08692$}\\
C\,{\sc i}            & 1328,1560,1656                & $12.82\pm 0.11$\phantom{$^{\rm a}$} & \phantom{0}$1.1\pm 0.3$\phantom{$^{\rm b}$}\\
C\,{\sc i}\,$^\star$  & 1656,1657.3,1657.9 $^{\rm a}$ & $12.42\pm 0.12$\phantom{$^{\rm a}$} & \phantom{uuu..}''                          \\
C\,{\sc ii}\,$^\star$ & 1335.6,1335.7                 & $12.78\pm 0.20$\phantom{$^{\rm a}$} & \phantom{uuu..}''                          \\
N\,{\sc i}   & 1134.1,1134.4,1134.9                   & $14.46\pm 0.05$\phantom{$^{\rm a}$} & \phantom{0}$5.1\pm 0.5$\phantom{$^{\rm b}$}\\
Al\,{\sc ii} & 1670                                   & $11.74\pm 0.10$\phantom{$^{\rm a}$} & \phantom{uuu..}''                          \\
Si\,{\sc ii} & 1304,1526,1808                         & $13.82\pm 0.06$\phantom{$^{\rm a}$} & \phantom{uuu..}''                          \\
P\,{\sc ii}  & 1152                                   & $12.60\pm 0.08$\phantom{$^{\rm a}$} & \phantom{uuu..}''                          \\
S\,{\sc ii}  & 1253,1259                              & $14.17\pm 0.06$\phantom{$^{\rm a}$} & \phantom{uuu..}''                          \\
Fe\,{\sc ii} & 1121,1143,1608                         & $13.23\pm 0.09$\phantom{$^{\rm a}$} & \phantom{uuu..}''                          \\
Zn\,{\sc ii} & 2026,2062                              & $11.77\pm 0.07$\phantom{$^{\rm a}$} & \phantom{uuu..}''                          \\
\hline
\end{tabular}
\label{tabmet3}
\flushleft Note: the two components of the C\,{\sc i},
C\,{\sc i}\,$^\star$ and C\,{\sc ii}\,$^\star$ profiles are redshifted
by $\sim 5$ km s$^{-1}$ compared to what is observed for the other metal
lines (whose redshifts are indicated).\\
$^{\rm a}$ Also 1329.08, 1329.10, 1329.12, 1560.6, 1560.7.\\
$^{\rm b}$ $5\sigma$ upper limit.
\end{table}

\begin{figure*}
\flushleft{\vbox{
\psfig{figure=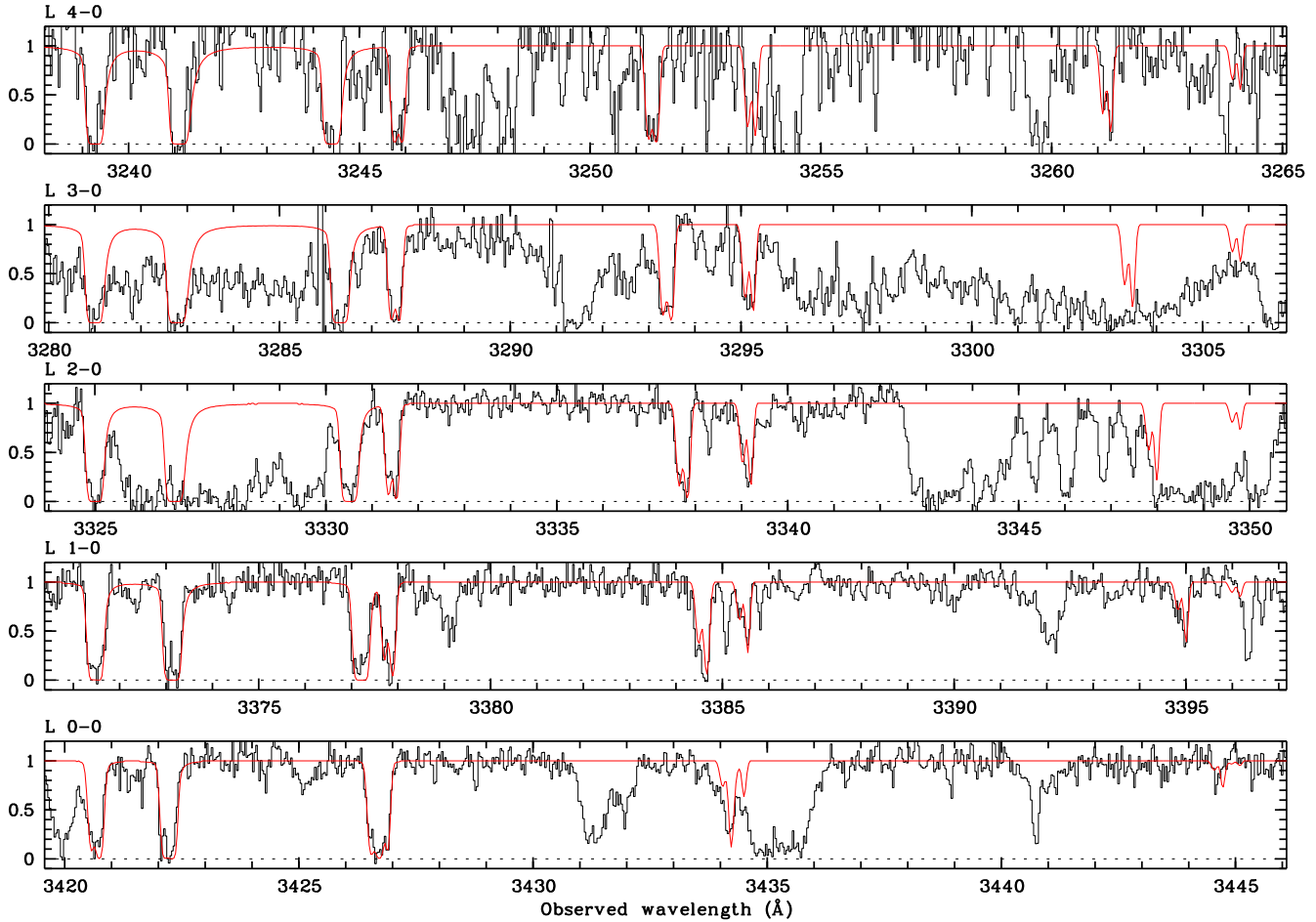,width=17.7cm,clip=,bbllx=41.pt,bblly=63.pt,bburx=555.pt,bbury=778.pt,angle=270.}}}
\caption[]{Voigt-profile fitting to the transition lines from the $J=0$, 1,
2, 3 and 4 rotational levels of the vibrational ground-state Lyman band of
H$_2$ at $z_{\rm abs}\approx 2.087$ toward Q\,1444$+$014. H$_2$ is detected on
this line of sight in two gas clouds spanning 12 km s$^{-1}$ in
velocity space.}
\label{figmole3}
\end{figure*}

Molecular hydrogen lines from the $J=0$, 1, 2 and 3 rotational levels are
detected in two different components most clearly seen in $J\ge 2$ (see
Fig.~\ref{figmole3}). Damping wings are definitively absent in $J=0$ and 1.
The redshifts of the H$_2$ components are consistent with those of
the C\,{\sc i} lines. Since the latter are accurately determined,
we simultaneously fitted all unblended H$_2$ lines from different
rotational levels with two components located at the measured redshifts of the
C\,{\sc i} lines. For a given component, the same broadening parameter was
used independently of the $J$ level of the lines. The column density in each
$J$ level was derived from several trials of Voigt-profile fitting to
take into account the range of possible $b$ values. Errors in the
column densities therefore correspond to a range of column densities and not
to the rms error from fitting the Voigt profiles. The results are given in
Table~\ref{tabmol3}.

The line profiles of S\,{\sc ii} and Zn\,{\sc ii} on the one hand, and
Si\,{\sc ii} and Fe\,{\sc ii} on the other, are different in the central clump
(see Fig.~\ref{figmetals3}). Whereas the S\,{\sc ii} and Zn\,{\sc ii}
profiles have a rectangular shape with two main components, the redder of
these two components is barely detected in the wing of the clump in both
Si\,{\sc ii}\,$\lambda$1304 and Fe\,{\sc ii}\,$\lambda$1608. This
demonstrates that Si and Fe depletions are much larger in the latter
component. The redshifts of the main components of the system, in addition
to a third one, were determined from simultaneous fitting of most of
the observed metal lines but independently of the fitting to
the C\,{\sc i} lines. This shows an apparent velocity offset of $\sim 5$
km s$^{-1}$ between C\,{\sc i} and other metal lines. The ionic
column densities in the component in the red wing of the clump could also
be derived accurately for a few species in the case of refractory
elements using several transition lines covering a range of oscillator
strengths (see Table~\ref{tabmet3}). We find [Zn/Si$]=0.41\pm 0.13$ and
[Zn/Fe$]=0.69\pm 0.13$ at $z_{\rm abs}=2.08679$, and
[Zn/Si$]=0.85\pm 0.09$ and [Zn/Fe$]=1.40\pm 0.11$ at
$z_{\rm abs}=2.08692$. This is reminiscent of warm and cold Galactic disc
cloud dust-depletion patterns respectively. The large depletion
factors measured in the component at $z_{\rm abs}=2.08692$ are of the
same order of magnitude as those measured in the component
at $z_{\rm abs}=1.96822$ toward Q\,0013$-$004 (Petitjean et al. 2002). This
shows that, even though the depletion factors averaged over the
entire profiles of DLA systems are usually moderate, they can be quite
large in individual components.

\section{The H$_2$-survey sample}\label{sam}

In this Section, we discuss all H$_2$ measurements performed to date at high
spectral resolution in DLA and sub-DLA systems at high
redshift ($z_{\rm abs}>1.8$). These measurements are summarized
in Table~\ref{met}. The H$_2$ and H\,{\sc i} column densities are the total
ones integrated over the systems, except for Q\,0013$-$004 for which two
sub-systems could be studied individually. In most of the systems, there
are several metal components and therefore the values for individual
components may be scattered around the quoted mean. For each of the
systems, we also give the Fe abundance, [Fe/H], and the metallicity, [X/H],
relative to Solar abundances. The metallicity was measured from the
abundance of Zn when Zn\,{\sc ii} is detected, and we used either S or
Si otherwise. As Zn is known to be little or undepleted onto dust grains,
the [Zn/Fe] ratio should be a good indicator of dust depletion. S itself
is undepleted onto dust grains. However, possible overabundance
of $\alpha$-elements relative to Fe at low metallicity may lead to
overestimate the dust-depletion factor when using [S/Fe]. However, the mean
overabundance of $\alpha$-elements should in general be less than about 0.25
dex (see Prochaska \& Wolfe 2002). Si can be depleted onto dust grains and,
hence, the ratio [Si/Fe] may lead to underestimate the dust-depletion
factor, but probably by less than 0.3 dex (see Petitjean et al. 2002). In
the following, however, we do not attempt to derive exact values of
the dust-depletion factor but only trends in the global behavior of
the characteristics of the whole population. It will be shown that these
considerations have little impact on the global results.

\begin{table}
\caption {Voigt-profile fitting results for different rotational levels of the
vibrational ground-state Lyman band of H$_2$ toward Q\,1444$+$014}
\begin{tabular}{llll}
\hline
\hline
$z_{\rm abs}$ & Level & $\log N\pm\sigma _{\log N}$ & $b\pm\sigma _b$\\
              &       & (H$_2$)                     & (km s$^{-1}$)  \\
\hline
2.08680 & $J=0$ & $15.68^{+0.22}_{-0.08}$ &  $5.5\pm 0.5$    \\
        & $J=1$ & $16.37^{+0.27}_{-0.11}$ &  \phantom{uu..}''\\
        & $J=2$ & $15.31^{+0.06}_{-0.03}$ &  \phantom{uu..}''\\
        & $J=3$ & $14.80^{+0.06}_{-0.04}$ &  \phantom{uu..}''\\
        & $J=4$ & $\le 14.15$ $^{\rm a}$  &  \phantom{uu..}''\\
        & $J=5$ & $<13.75$ $^{\rm b}$     &  \phantom{uu..}''\\
2.08696 & $J=0$ & $17.46^{+0.10}_{-0.11}$ &  $2.6\pm 0.6$    \\
        & $J=1$ & $18.03^{+0.12}_{-0.12}$ &  \phantom{uu..}''\\
        & $J=2$ & $16.63^{+0.57}_{-0.46}$ &  \phantom{uu..}''\\
        & $J=3$ & $15.26^{+0.07}_{-0.08}$ &  \phantom{uu..}''\\
        & $J=4$ & $\le 14.20$ $^{\rm a}$  &  \phantom{uu..}''\\
        & $J=5$ & $<13.75$ $^{\rm b}$     &  \phantom{uu..}''\\
%2.08680 & $J=0$ &     $15.83\pm 0.20$ &  $5.5\pm 0.3$  \\
%        & $J=1$ &     $16.45\pm 0.15$ &   ''           \\
%        & $J=2$ &     $15.29\pm 0.02$ &   ''           \\
%        & $J=3$ &     $14.78\pm 0.03$ &   ''           \\
%        & $J=4$ & $\le 14.13\pm 0.05$ &   ''           \\
%        & $J=5$ &    $<13.53^{\rm a}$ &   ''           \\
%2.08696 & $J=0$ &     $17.72\pm 0.45$ &  $3.2\pm 0.2$  \\
%        & $J=1$ &     $18.15\pm 0.28$ &   ''           \\
%        & $J=2$ &     $16.28\pm 0.14$ &   ''           \\
%        & $J=3$ &     $15.19\pm 0.03$ &   ''           \\
%        & $J=4$ & $\le 14.20\pm 0.04$ &   ''           \\
%        & $J=5$ &    $<13.53^{\rm a}$ &   ''           \\
\hline
\end{tabular}
\label{tabmol3}
\flushleft Note: errors in the column densities correspond to a range
of column densities; they are not the rms errors from fitting the
Voigt profiles (see text).\\
$^{\rm a}$ Possible blends.\\
$^{\rm b}$ $5\sigma$ upper limit.
\end{table}

\begin{figure}
\centerline{\hbox{
\psfig{figure=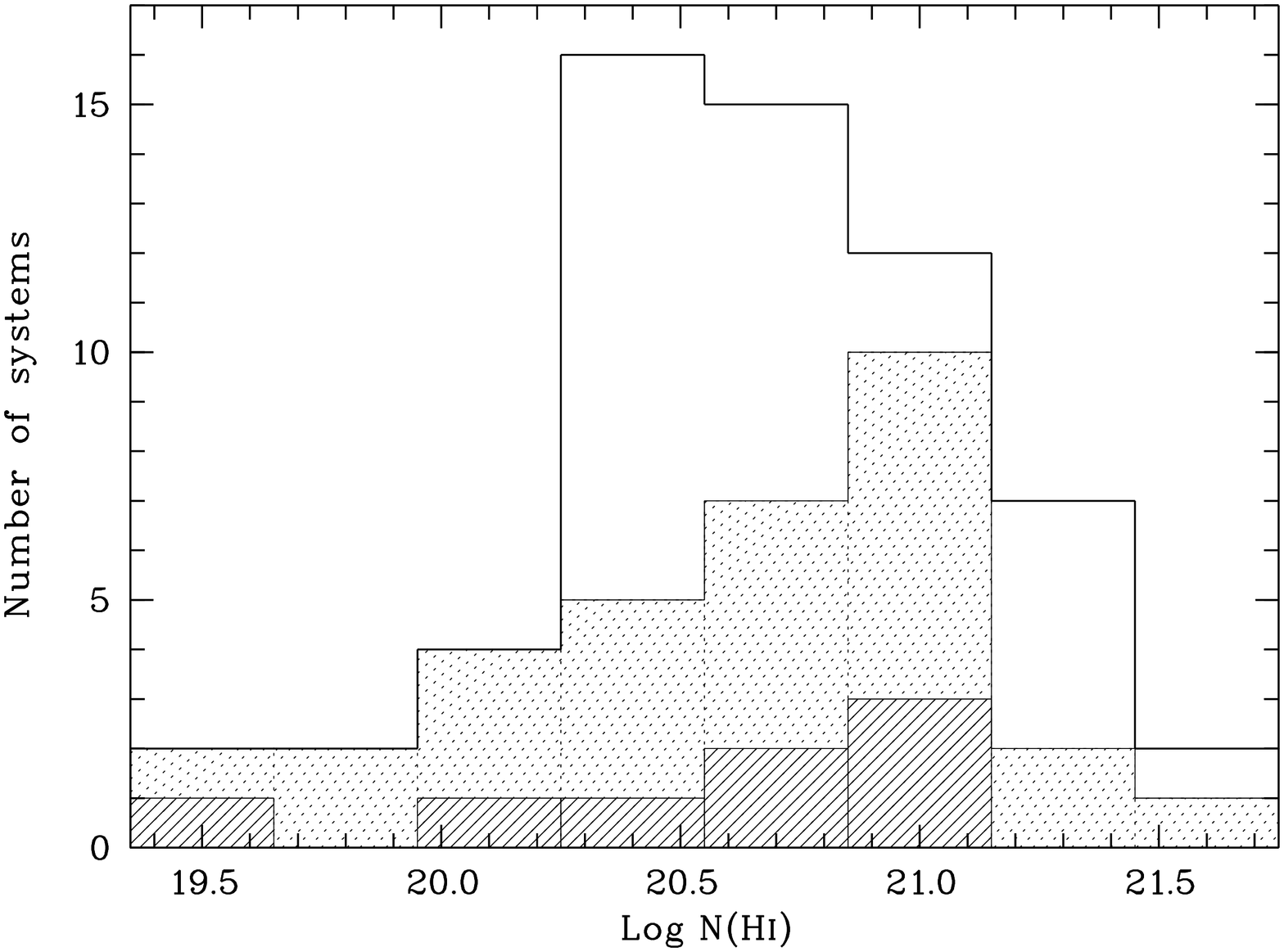,width=8.4cm,clip=,bbllx=42.pt,bblly=74.pt,bburx=569.pt,bbury=785.pt,angle=270.}}}
\centerline{\hbox{
\psfig{figure=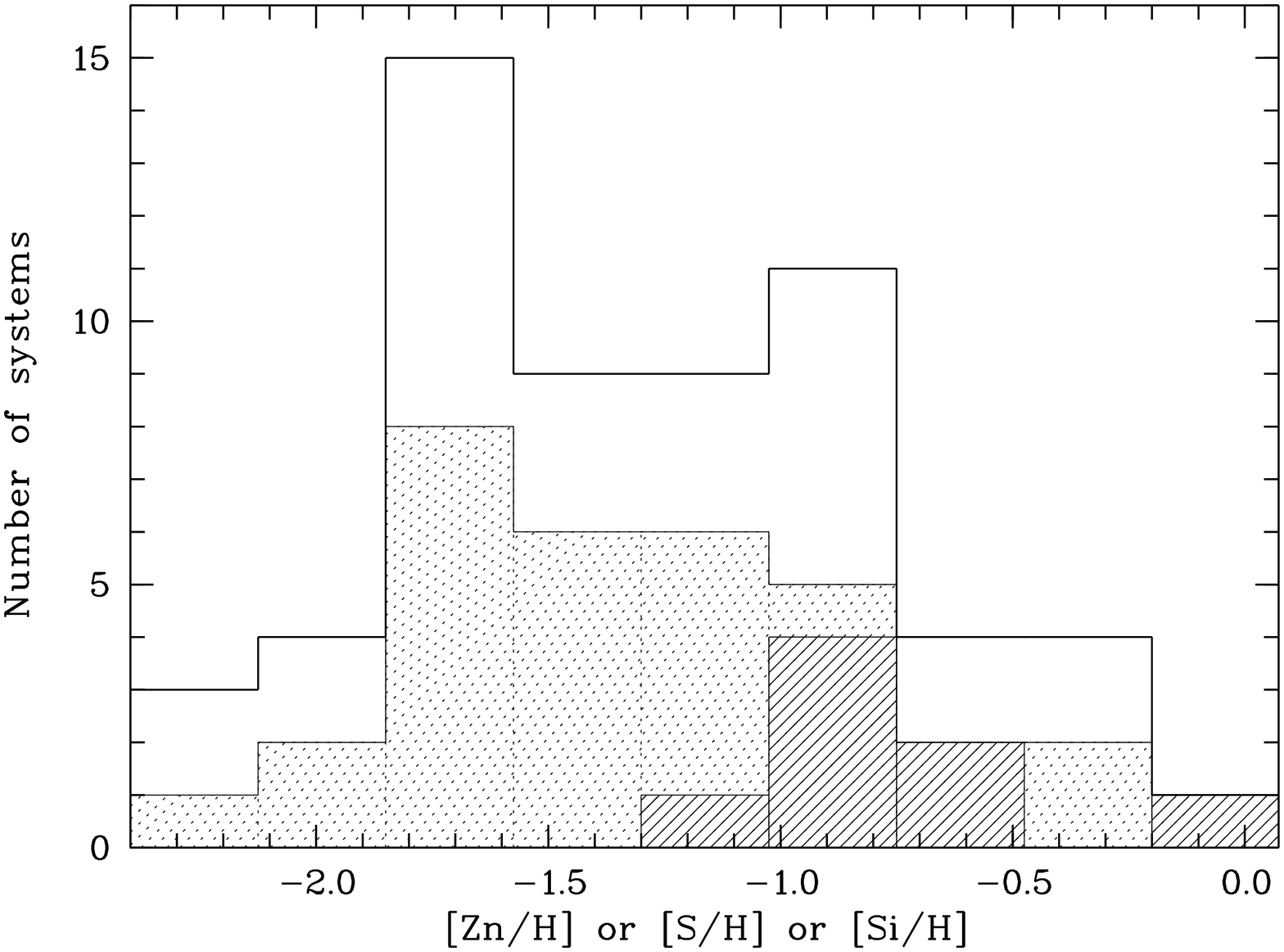,width=8.4cm,clip=,bbllx=42.pt,bblly=74.pt,bburx=569.pt,bbury=785.pt,angle=270.}}}
\centerline{\hbox{
\psfig{figure=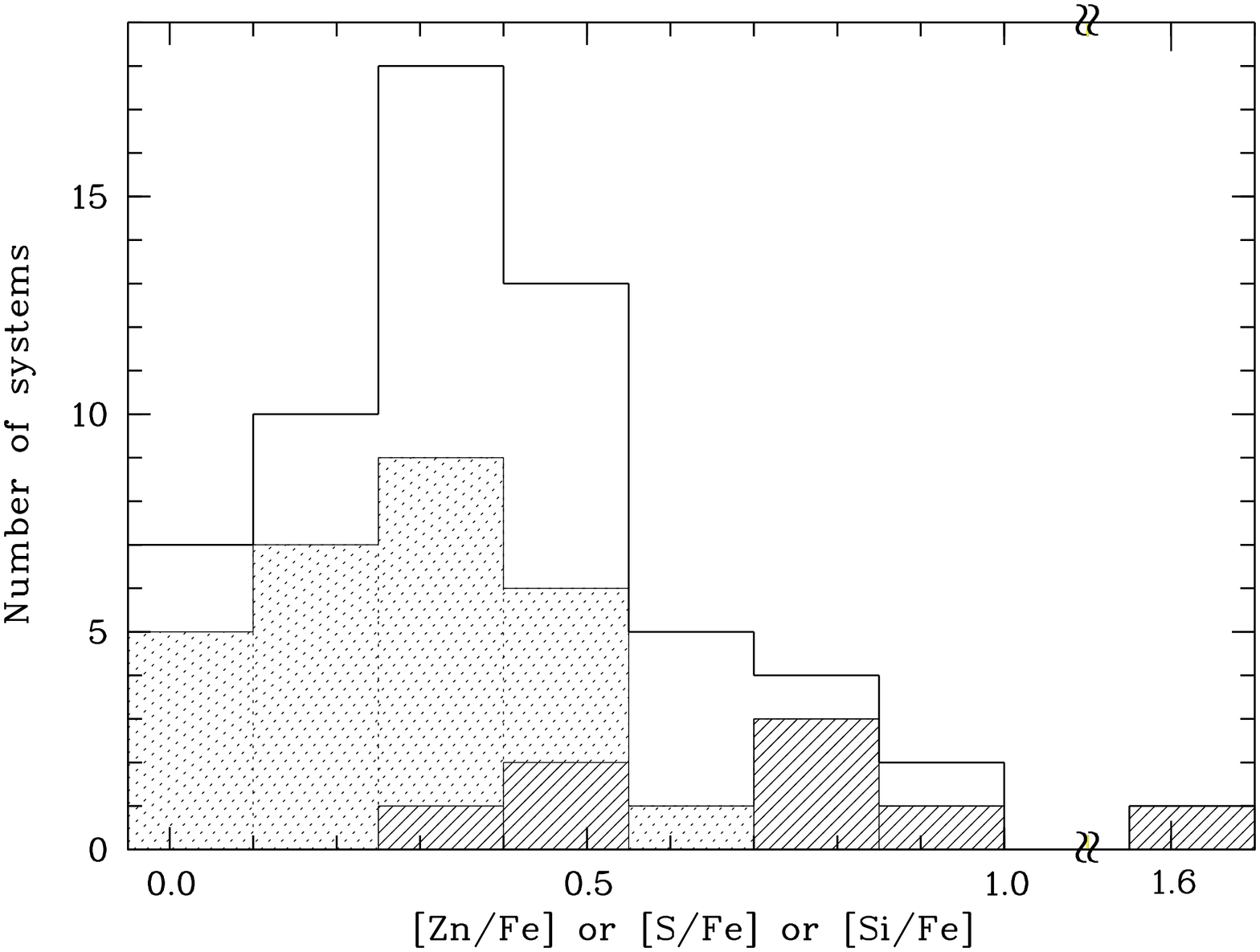,width=8.4cm,clip=,bbllx=42.pt,bblly=74.pt,bburx=569.pt,bbury=785.pt,angle=270.}}}
\caption[]{Distributions of neutral hydrogen column densities
({\sl upper panel}), metallicities ({\sl middle panel}; measured from either
[Zn/H], or [S/H], or [Si/H]) and depletion factors ({\sl lower panel};
measured from either [Zn/Fe], or [S/Fe], or [Si/Fe]) in, respectively, the
global population of DLA/sub-DLA systems (overall distribution), our
H$_2$-survey sample (dotted histogram) and the sub-sample of H$_2$-detected
systems (hashed histogram).}
\label{figsample1}
\end{figure}

\begin{figure}
\flushleft{\vbox{
\psfig{figure=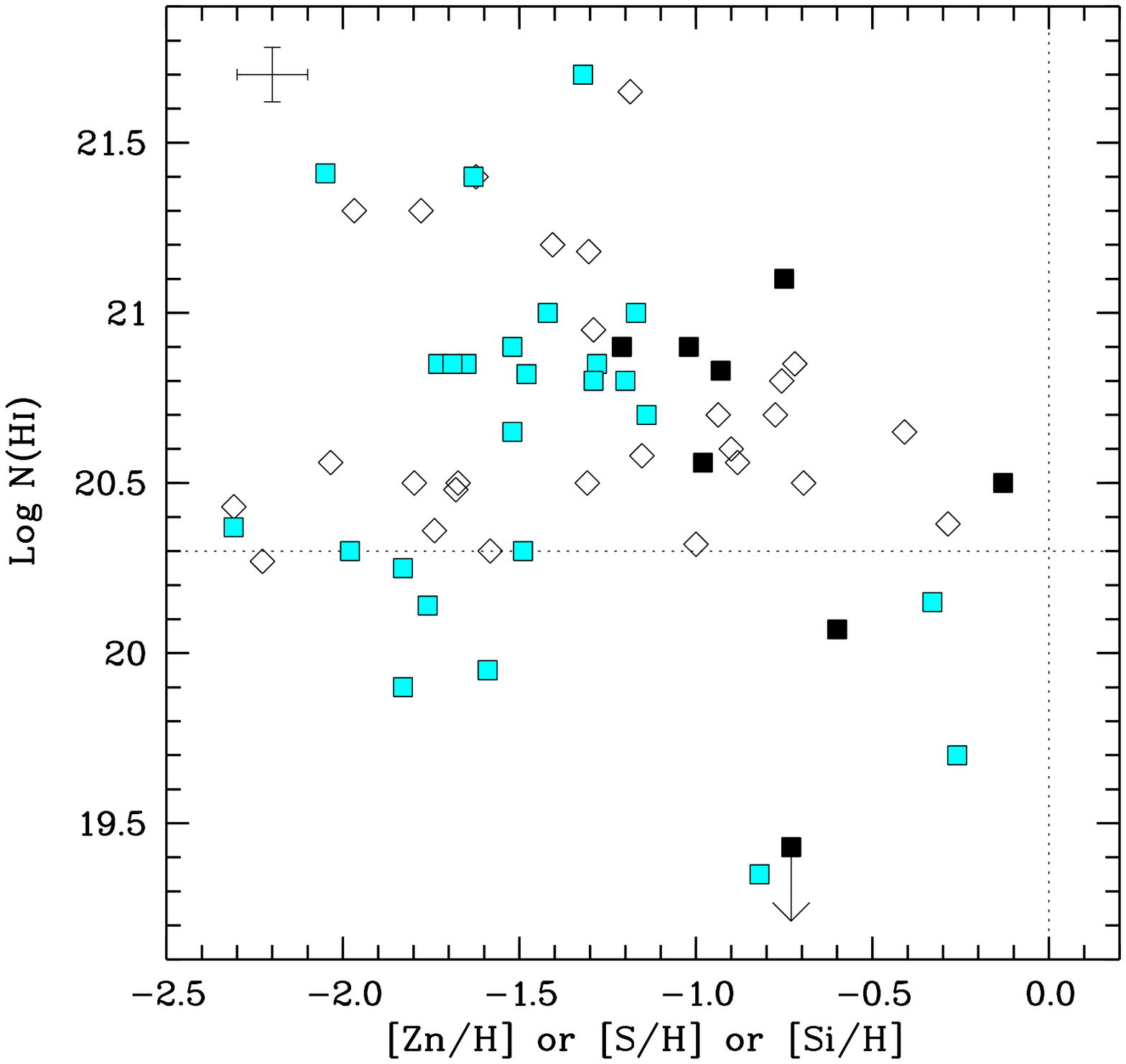,width=8.4cm,clip=,bbllx=35.pt,bblly=291.pt,bburx=550.pt,bbury=772.pt,angle=0.}}}
\vspace{-0.83cm}\hspace{-0.165cm}
\centerline{\hbox{
\psfig{figure=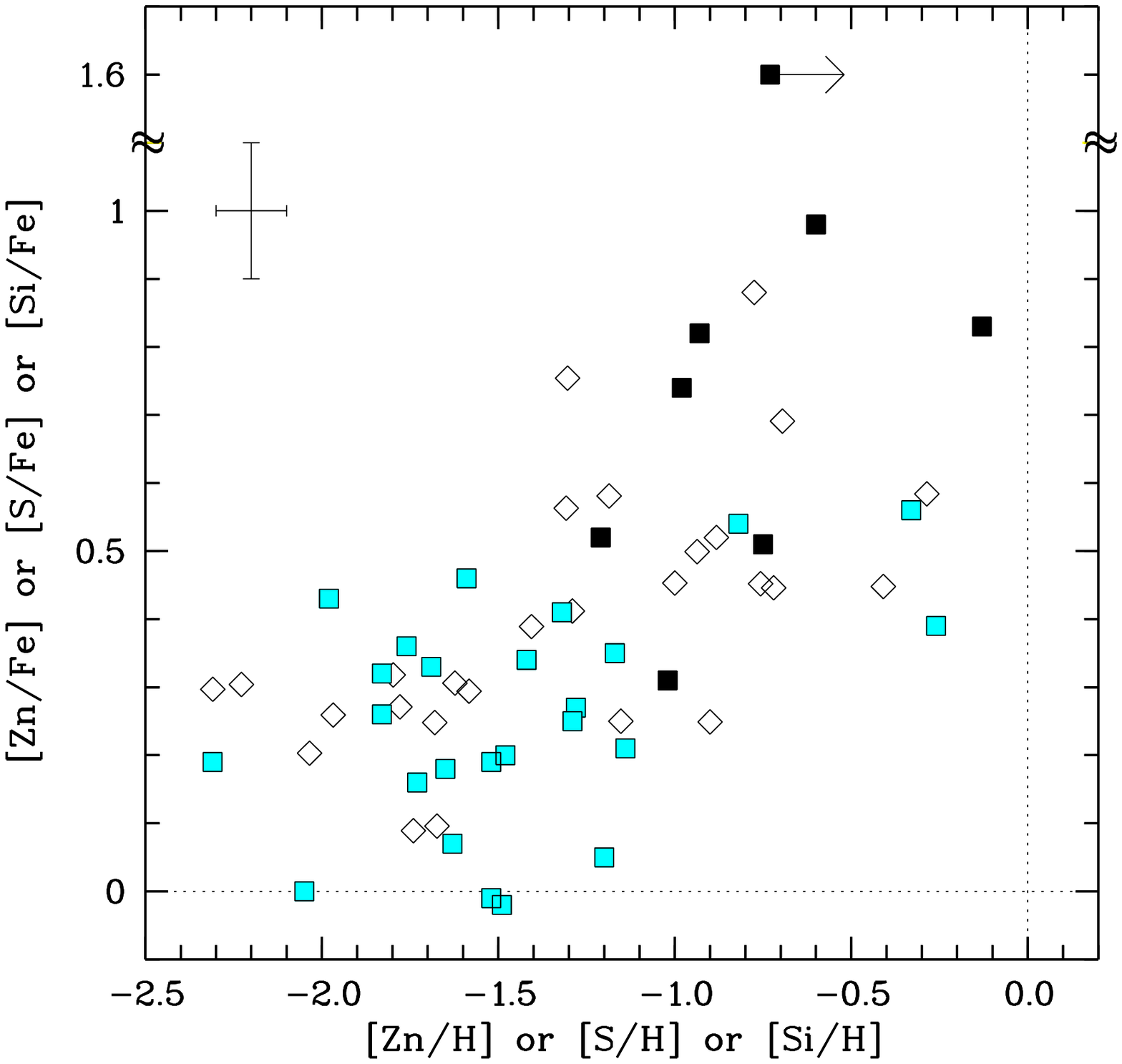,width=8.4cm,clip=,bbllx=35.pt,bblly=291.pt,bburx=550.pt,bbury=772.pt,angle=0.}}}
\caption[]{Neutral hydrogen column density ({\sl upper plot}) and depletion
factor ({\sl lower plot}) versus metallicity in, respectively, the
global population of DLA/sub-DLA systems (all symbols), our H$_2$-survey sample
(all squares) and the sub-sample of H$_2$-detected systems (dark squares). The
typical error bars are shown in the upper-left part of each plot. A
correlation between metallicity and depletion factor is present at the
$4\sigma$ significance level, confirming the trend previously detected by
Ledoux et al. (2002a).}
\label{figsample2}
\end{figure}

In Figs.~\ref{figsample1} and \ref{figsample2}, we compare the
characteristics of our sample of 33 systems (called in the following sample
S1) with those of the global population of DLA systems. For the latter, we use
the result of merging our sample with the sample of Prochaska et al. (2001).
The corresponding sample will be called sample Spop and comprises 60
systems. In Fig.~\ref{figsample1}, different distributions for the
global population of DLA/sub-DLA systems (sample Spop), sample S1 and
the sub-sample of systems where H$_2$ is detected (sample SH$_2$) are shown.
%respectively, with a continuous line (overall distribution), as a dotted
%histogram and as a hashed histogram.

%New table
\begin{table*}
\caption{Summary of molecular and metal contents in the H$_2$-survey sample
of DLA/sub-DLA systems}
\begin{tabular}{llllllllll}
\hline
\hline
QSO & $z_{\rm em}$ & $z_{\rm abs}$ & $\log N($H\,{\sc i}$)$ & \multicolumn{2}{c}{$\log N($H$_2)$ $^1$} & $\log f$ $^2$ & [Fe/H$]$ $^3$ & [X/H$]$ $^3$ &  \\
    &              &               &                        & $J=0$ & $J=1$                            &               &               &              & X\\
\hline
0000$-$263 & 4.10 & 3.390 & $21.41\pm 0.08$ $^{\rm a}$ & $<13.9$            & $13.74$:: $^{\rm b}$                      & $<-6.98$              & $-2.05\pm 0.09$ $^{\rm c}$ & $-2.05\pm 0.09$ $^{\rm c}$ & Zn\\
0010$-$002 & 2.14 & 2.025 & $20.80\pm 0.10$            & $<14.0$            & $<14.2$                                   & $<-6.09$              & $-1.25\pm 0.11$            & $-1.20\pm 0.12$            & Zn\\
0013$-$004 & 2.09 & 1.968 & $\le 19.43$ $^{\rm d}$     & \multicolumn{2}{l}{\phantom{aaaa.}$16.77^{+0.05}_{-0.07}$ $^{\rm d}$} & $\ge -2.36$ $^{\rm d}$       & $\ge -2.33$ $^{{\rm d},4}$       & $\ge -0.73$ $^{{\rm d},4}$ & Zn\\
0013$-$004 & 2.09 & 1.973 & $20.83\pm 0.05$ $^{\rm d}$ & \multicolumn{2}{l}{\phantom{aaaa.}$17.72$ / $20.00$ $^{\rm d}$}       & $-2.81$ / $-0.64$ $^{\rm d}$ & $-1.75\pm 0.05$ $^{\rm d}$       & $-0.93\pm 0.06$ $^{\rm d}$ & Zn\\
0058$-$292 & 3.09 & 2.671 & $21.00\pm 0.10$            & $<13.8$            & $<13.6$                                   & $<-6.69$              & $-1.76\pm 0.10$            & $-1.42\pm 0.11$            & Zn\\
0102$-$190 & 3.04 & 2.370 & $20.85\pm 0.08$            & $<14.7$            & $<14.6$                                   & $<-5.60$              & $-1.89\pm 0.13$            & $-1.73\pm 0.14$            & Zn\\
0112$-$306 & 2.98 & 2.418 & $20.37\pm 0.08$            & $<13.8$            & $<14.2$                                   & $<-5.72$              & $-2.50\pm 0.09$            & $-2.31\pm 0.08$            & Si\\
0112$-$306 & 2.98 & 2.702 & $20.15\pm 0.07$            & $<14.1$            & $<14.0$                                   & $<-5.50$              & $-0.89\pm 0.10$            & $-0.33\pm 0.11$            & Si\\
0112$+$029 & 2.81 & 2.423 & $20.70\pm 0.10$            & $<13.6$            & $<13.9$                                   & $<-6.32$              & $-1.35\pm 0.11$            & $-1.14\pm 0.13$            & S \\
0135$-$273 & 3.21 & 2.800 & $20.80\pm 0.10$            & $<13.8$            & $<13.8$                                   & $<-6.40$              & $-1.54\pm 0.15$            & $-1.29\pm 0.17$            & S \\
0347$-$383 & 3.22 & 3.025 & $20.56\pm 0.05$            & \multicolumn{2}{l}{\phantom{aaaa.}$14.55\pm 0.09$}             & $-5.71\pm 0.10$       & $-1.72\pm 0.06$            & $-0.98\pm 0.09$            & Zn\\
0405$-$443 & 3.02 & 2.550 & $21.00\pm 0.15$            & $<13.9$            & $<13.7$                                   & $<-6.59$              & $-1.52\pm 0.15$            & $-1.17\pm 0.16$            & Zn\\
0405$-$443 & 3.02 & 2.595 & $20.90\pm 0.10$            & \multicolumn{2}{l}{\phantom{aaaa.}$18.16^{+0.21}_{-0.06}$} & $-2.44^{+0.23}_{-0.12}$   & $-1.33\pm 0.11$            & $-1.02\pm 0.12$            & Zn\\
0405$-$443 & 3.02 & 2.621 & $20.25\pm 0.10$            & $<13.5$            & $<13.5$                                   & $<-6.15$              & $-2.15\pm 0.10$            & $-1.83\pm 0.10$            & Si\\
0528$-$250 & 2.78 & 2.811 & $21.10\pm 0.10$            & \multicolumn{2}{l}{\phantom{aaaa.}$18.22^{+0.23}_{-0.16}$ $^{\rm 5}$} & $-2.58^{+0.25}_{-0.19}$ $^{\rm 5}$ & $-1.26\pm 0.10$            & $-0.75\pm 0.10$            & Zn\\
0551$-$366 & 2.32 & 1.962 & $20.50\pm 0.08$ $^{\rm e}$ & \multicolumn{2}{l}{\phantom{aaaa.}$17.42^{+0.63}_{-0.90}$ $^{\rm e}$} & $-2.78^{+0.64}_{-0.90}$ $^{\rm e}$ & $-0.96\pm 0.09$ $^{\rm e}$ & $-0.13\pm 0.09$ $^{\rm e}$ & Zn\\
0841$+$129 & 2.50 & 2.374 & $20.90\pm 0.10$            & $14.56$::          & $<14.0$                                   & $<-5.93$              & $-1.71\pm 0.11$            & $-1.52\pm 0.14$            & Zn\\
0841$+$129 & 2.50 & 2.476 & $20.65\pm 0.10$            & $<14.0$            & $<14.0$                                   & $<-6.05$              & $-1.51\pm 0.11$            & $-1.52\pm 0.11$            & S \\
1037$-$270 & 2.23 & 2.139 & $19.70\pm 0.10$ $^{\rm f}$ & $<13.7$ $^{\rm f}$ & $<13.7$ $^{\rm f}$                        & $<-5.40$ $^{\rm f}$   & $-0.65\pm 0.10$ $^{\rm f}$ & $-0.26\pm 0.11$ $^{\rm f}$ & Zn\\
1101$-$264 & 2.14 & 1.839 & $19.35\pm 0.04$            & $<14.0$            & $<14.0$                                   & $<-4.75$              & $-1.36\pm 0.05$            & $-0.82\pm 0.14$            & S \\
1117$-$133 & 3.96 & 3.351 & $20.85\pm 0.10$ $^{\rm g}$ & $<13.7$            & $<14.0$                                   & $<-6.37$              & $-1.55\pm 0.13$ $^{\rm g}$ & $-1.28\pm 0.13$ $^{\rm g}$ & Zn\\
1157$+$014 & 1.99 & 1.944 & $21.70\pm 0.10$            & $<14.3$            & $<14.5$                                   & $<-6.69$              & $-1.73\pm 0.10$            & $-1.32\pm 0.10$            & Zn\\
1223$+$178 & 2.94 & 2.466 & $21.40\pm 0.10$            & $<14.0$            & $<14.0$                                   & $<-6.80$              & $-1.70\pm 0.10$            & $-1.63\pm 0.11$            & Zn\\
1232$+$082 & 2.58 & 2.338 & $20.90\pm 0.10$ $^{\rm h}$ & \multicolumn{2}{l}{\phantom{aaaa.}$\ge 17.19$ $^{\rm h,6}$} & $\ge -3.41$ $^{\rm h,6}$ & $-1.73\pm 0.13$ $^{\rm h}$ & $-1.21\pm 0.15$ $^{\rm h}$ & Si\\
1337$+$113 & 2.92 & 2.508 & $19.95\pm 0.05$            & $<13.8$            & $<14.1$                                   & $<-5.37$              & $-2.05\pm 0.07$            & $-1.59\pm 0.09$            & Si\\
1337$+$113 & 2.92 & 2.796 & $20.85\pm 0.08$            & $<13.6$            & $<13.8$                                   & $<-6.54$              & $-2.02\pm 0.09$            & $-1.69\pm 0.11$            & Si\\
1444$+$014 & 2.21 & 2.087 & $20.07\pm 0.07$            & \multicolumn{2}{l}{\phantom{aaaa.}$18.30^{+0.37}_{-0.37}$}  & $-1.48^{+0.38}_{-0.38}$  & $-1.58\pm 0.09$            & $-0.60\pm 0.15$            & Zn\\
1451$+$123 & 3.25 & 2.469 & $20.30\pm 0.10$            & $<15.0$:           & $<15.0$:                                  & $<-4.70$              & $-2.41\pm 0.10$            & $-1.98\pm 0.14$            & Si\\
1451$+$123 & 3.25 & 3.171 & $19.90\pm 0.20$            & $<13.5$            & $<13.5$                                   & $<-5.80$              & $-2.09\pm 0.24$            & $-1.83\pm 0.21$            & Si\\
2059$-$360 & 3.09 & 2.508 & $20.14\pm 0.07$            & $<14.7$            & $<14.5$                                   & $<-4.93$              & $-2.12\pm 0.08$            & $-1.76\pm 0.09$            & S \\
2059$-$360 & 3.09 & 3.083 & $20.85\pm 0.08$            & $<13.5$            & $<13.7$                                   & $<-6.64$              & $-1.83\pm 0.09$            & $-1.65\pm 0.12$            & S \\
%2138$-$444 & 3.17 & 2.383 & $20.40\pm 0.05$            &                    &                                           &                       & $-0.99$                    & $-1.590$                   &   \\
2138$-$444 & 3.17 & 2.852 & $20.82\pm 0.05$            & $<13.5$            & $<13.1$                                   & $<-6.87$              & $-1.68\pm 0.05$            & $-1.48\pm 0.05$            & Zn\\
2332$-$094 & 3.30 & 3.057 & $20.30\pm 0.08$            & $<12.8$            & $<13.1$                                   & $<-6.72$              & $-1.47\pm 0.08$            & $-1.49\pm 0.20$            & S \\
\hline
\end{tabular}
\label{met}
\flushleft $^1$ We give total molecular hydrogen column densities summed
up over all $J$ levels in case of detection and upper limits for $J=0$ and
1 in case of non-detection.\\
$^2$ Molecular fraction $f=2N($H$_2)/(2N($H$_2)+N($H\,{\sc i}$))$.\\
$^3$ Fe abundances, [Fe/H], and metallicities, [X/H] with either
${\rm X}={\rm Zn}$, or S, or Si. In order to derive meaningful abundance
ratios, the components taken into account in the profiles are
those simultaneously detected in the weak lines of Zn\,{\sc ii}, S\,{\sc ii}
and Si\,{\sc ii} (see Ledoux et al. 2002a).\\
$^4$ [Zn/Fe$]=1.60\pm 0.04$.\\
$^5$ The fitting was performed on the new UVES data taking into account
the two-component structure of the cloud.\\
$^6$ The total H$_2$ column density of this system will be derived from higher
quality data. The value derived by Srianand et al. (2000) should be
considered as a lower limit.\\
{\sc References}: hydrogen column density and metal abundance measurements
are from this work, unless otherwise indicated:
(a)~Lu et al. (1996);
(b)~Levshakov et al. (2001);
(c)~Molaro et al. (2001);
(d)~Petitjean et al. (2002);
(e)~Ledoux et al. (2002b);
(f)~Srianand \& Petitjean (2001);
(g)~P\'eroux et al. (2002);
(h)~Srianand et al. (2000).
\end{table*}

The neutral hydrogen column density distributions are shown in the upper panel
of Fig.~\ref{figsample1}. There are 28 (resp. 32) systems with
$\log N($H\,{\sc i}$)$ smaller (resp. larger) than 20.6 in sample Spop, out
of which 50 (resp. 59) percent belongs to sample S1. The two samples
are statistically indistinguishable although the mean H\,{\sc i}
column density is slightly smaller in our sample. The Kolmogorov--Smirnov test
probability that the two distributions are drawn from the same
parent population is $P_{\rm KS}=0.91$ (two-sided case). It is also apparent,
although the statistics are based on a smaller data set, that there is no
systematic correlation between the detection of H$_2$ lines and
the H\,{\sc i} column density. The $\log N($H\,{\sc i}$)$ distributions are
similar in sub-sample SH$_2$ and in sample S1 ($P_{\rm KS}>0.9999$).

It can be seen in the middle panel of Fig.~\ref{figsample1} that
the metallicity distributions are similar in sample S1 and in the
global population Spop ($P_{\rm KS}>0.9999$). There are 31 (resp. 29) systems
with [X/H] smaller (resp. larger) than $-1.3$ in sample Spop, out of which 55
(resp. 55) percent are part of sample S1. The mean metallicity is slightly
smaller in our sample however. This could be a consequence of our
sample including a large number of lines of sight with several DLA
systems. Indeed, if these systems had a high metallicity and therefore a
large dust content, they would have obscured the background quasars.
However, this does not mean that these lines of sight are biased against the
presence of H$_2$. We indeed detect H$_2$ at $z_{\rm abs}=2.595$ toward
Q\,0405$-$443, which is a line of sight containing no less than three DLA
systems. The important result is that the [X/H] distributions are different
in sub-sample SH$_2$ and in sample S1 ($P_{\rm KS}=0.03$): systems
where H$_2$ is detected are definitively amongst the most metal-rich.
Interestingly, none of the 17 systems with [X/H$]<-1.3$ in sample S1
show detectable amount of H$_2$. However, H$_2$ is detected down
to [X/H$]\approx -1.2$.

The distributions of depletion factors (Fig.~\ref{figsample1}, lower panel) are
also not significantly different in sample S1 compared to
sample Spop ($P_{\rm KS}=0.98$). There are 35 (resp. 25) systems with [X/Fe]
smaller (resp. larger) than 0.4 in sample Spop, out of which 60 (resp.
48) percent belongs to sample S1. Although H$_2$ is detected in a DLA system
with [Zn/Fe] as low as 0.3 (at $z_{\rm abs}=2.595$ toward Q\,0405$-$443),
there is a clear tendency for depletion factors to be larger in systems where
H$_2$ is detected. The [X/Fe] distributions are different in sub-sample SH$_2$
and in sample S1 ($P_{\rm KS}=0.02$). In particular, H$_2$ is detected in
the five systems with the largest depletion factors ([X/Fe$]\ge 0.7$).

\begin{figure*}
\centerline{\vbox{
\psfig{figure=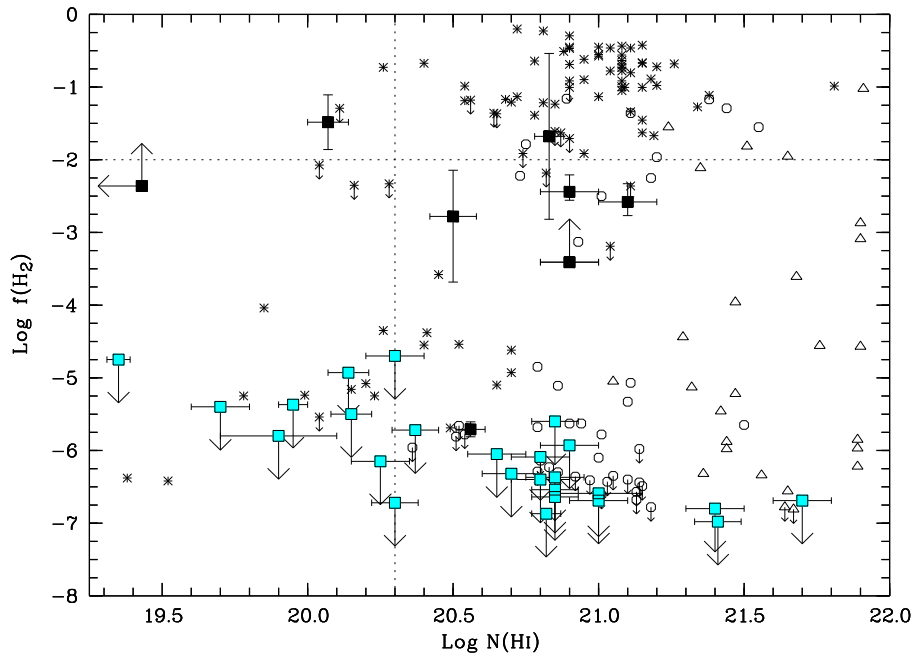,width=12.cm,clip=,bbllx=47.pt,bblly=82.pt,bburx=560.pt,bbury=788.pt,angle=270.}}}
\caption[]{Mean H$_2$ molecular
fraction $f=2N($H$_2)/(2N($H$_2)+N($H\,{\sc i}$))$ versus neutral hydrogen
column density. Measurements in DLA/sub-DLA systems are indicated by dark
squares for H$_2$ detections and shaded ones for upper limits. Observations
along lines of sight in the Galaxy (Savage et al. 1977) and the LMC and SMC
(Tumlinson et al. 2002) are indicated by, respectively, asterisks, circles
and triangles.}
\label{fh1}
\end{figure*}

In Fig.~\ref{figsample2}, the H\,{\sc i} column density (upper panel) and the
depletion factor (lower panel) are both plotted versus
metallicity. Measurements from sample S1 are shown by squares, with dark
squares for systems where H$_2$ is detected. It is apparent that, in the
overall population of DLA systems, there is a lack of systems with both a
high metallicity and a large H\,{\sc i} column density. This was first
noticed by Boiss\'e et al. (1998) who suggested that this might be due to
the fact that, if these systems exist, they may avoid detection because
the background quasars are attenuated by dust extinction (Fall \& Pei 1993).
This assumes at least some correlation between metallicity and the amount of
dust. It can be seen in the lower panel of Fig.~\ref{figsample2} that indeed
there is a trend for the most metal-rich systems to have larger depletion
factors. This is detected at the $4\sigma$ significance level using a
Kendall rank correlation test on censored data and taking into account the
measurement uncertainties. This trend was already noticed before
by Ledoux et al. (2002a), at a lower significance level, probably due to a
less homogenous sample, larger measurement uncertainties and the
observation of a narrower metallicity range. However, from this alone, it was
unclear whether this trend is a consequence of dust depletion effects or
peculiar nucleosynthesis history of zinc (Prochaska \& Wolfe 2002). However,
a correlation is observed in both our sample and also the global population of
DLA systems. Moreover, it is clear from Fig.~\ref{figsample2} that H$_2$
is detected at both the highest metallicities and the largest
depletion factors. This strongly favors the fact that the correlation
between metallicity and depletion factor is a consequence of dust
depletion effects.

\section{Discussion}\label{res}

\subsection{Molecular hydrogen in DLA systems}

If we consider only DLA systems, with $\log N($H\,{\sc i}$)\ge 20.3$,
firm detection of H$_2$ is achieved in 6 out of 24 systems. If we also
include sub-DLA systems, this amounts to 8 detections out of 33
systems. Therefore, H$_2$ is detected in more than 20 percent of the DLA
systems. However, three of the systems where H$_2$ is seen were already known.
If we exclude these systems, we actually detect H$_2$ in 13 to 20 percent of
the newly surveyed systems.
Note that we do not include in these calculations the tentative detections
toward Q\,0000$-$263 (Levshakov et al. 2000) and Q\,0841$+$129
(Petitjean et al. 2000).

The detection probability of H$_2$ with column densities greater
than $10^{14}$ cm$^{-2}$ is more than 90 percent along lines of sight
through the ISM of our Galaxy and in the SMC (Savage et al. 1977;
Richter et al. 2001; Tumlinson et al. 2002). However, only 52 percent of the
lines of sight through the LMC have detectable H$_2$
lines (Tumlinson et al. 2002). These differences in the detection
probability cannot be ascribed solely to different chemical
enrichment histories as the ISM of the LMC has a mean metallicity only 0.3
dex lower than the Galactic local ISM and the metallicity of the SMC is
0.6 dex lower than that (Russell \& Dopita 1992). Therefore, the small
detection probability in DLA systems is probably not only related to their
low metallicities. The level of star-formation activity in or nearby the
systems might well be the primary reason for the low detection rate
of H$_2$ as it has been advocated in the case of the LMC (Kim et al. 1999).

As noted above, in our sample the detection of H$_2$ is not correlated
with the H\,{\sc i} column density. The H\,{\sc i} column density distribution
is similar in sample S1 and in the sub-sample of systems where H$_2$
is detected (see Fig.~\ref{figsample1}). In particular, H$_2$ is detected in
two sub-DLA systems, with $\log N($H\,{\sc i}$)<20.3$. In addition,
amongst the systems where H$_2$ is detected there is no correlation
between the logarithm of the molecular fraction, $\log f$, and
$\log N($H\,{\sc i}$)$ (see Fig.~\ref{fh1}).

In Fig.~\ref{fh1}, confirmed detections and upper limits on the
molecular fraction in sample S1 are represented by, respectively, dark and
shaded squares. Observations along lines of sight in the LMC and SMC
from Tumlinson et al. (2002) and in the Galaxy from Savage et al. (1977)
are also shown.
It is apparent that most DLA systems are similar to the LMC and/or SMC
lines of sight. In particular, and contrary to what happens in our Galaxy, a
significant number of lines of sight through the LMC have $\log f<-6$ for
$20.5\le\log N($H\,{\sc i}$)\le 21.5$. This is also the case for 58 percent
of the DLA systems ($20.3\le\log N($H\,{\sc i}$)\le 21.7$) in which we could
achieve such a limit, which means that this number is a lower limit on
the fraction of systems having such a small molecular fraction. Indeed, if
all detection limits are considered to be that constraining then the fraction
of systems with $\log f<-6$ is 75 percent. Note that several systems have
a very small molecular fraction, $\log f\la -7$. Interestingly, this is an
order of magnitude smaller than the primordial freeze-out fraction
of H$_2$ molecules (Lepp \& Shull 1984). It is very difficult to reproduce
such small values in models of cool clouds
(see Petitjean et al. 2000; Liszt 2002).
When H$_2$ is detected, the molecular fraction is in the
range $-3<\log f<-1$ with some scatter, with the exception of the DLA
system toward Q\,0347$-$383 for which $\log f=-5.71$. The former values
are similar to what is seen along the LMC and SMC lines of sight.

%*****************************************************************************
%
%NOTE FOR THE EDITOR: THE FOLLOWING TWO FIGURES SHOULD BE DISPLAYED AT THE TOP
%OF THE SAME PAGE AND SIDE BY SIDE
%
%*****************************************************************************
%
\begin{figure}
\flushleft{\vbox{
\psfig{figure=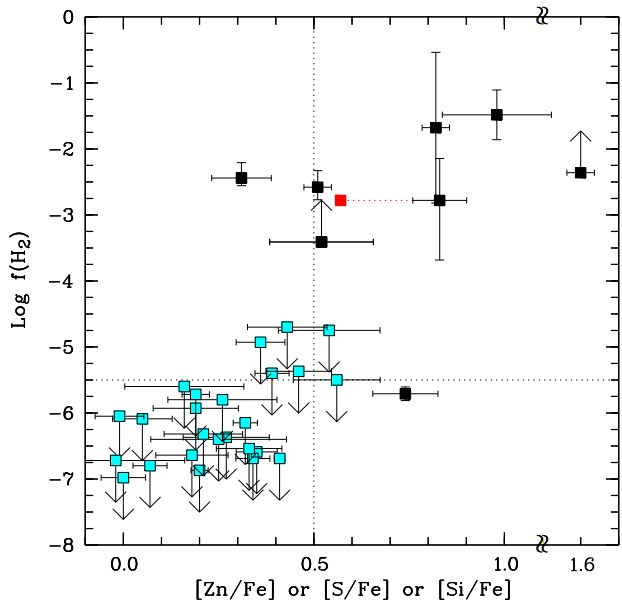,width=8.4cm,clip=,bbllx=42.pt,bblly=290.pt,bburx=554.pt,bbury=781.pt,angle=0.}}}
\caption[]{Mean H$_2$ molecular fraction versus depletion onto dust grains
as estimated from the abundance of either X$=$Zn, or S, or Si relative to Fe,
in the H$_2$-survey sample (see Table~\ref{met}). H$_2$ detections,
measured by our group, are indicated by dark squares. The dotted line
segment links data points for the DLA system toward Q\,0551$-$366 considering
either [Fe/Zn] or [Fe/S]. This is indeed the only one case where
the difference between measured [X/Fe] ratios is significantly larger than 0.1
dex (i.e. 0.24 dex). The two dotted lines are drawn for
illustrative purposes (see text).}
\label{figfdep}
\end{figure}
\begin{figure}
\flushleft{\vbox{
\psfig{figure=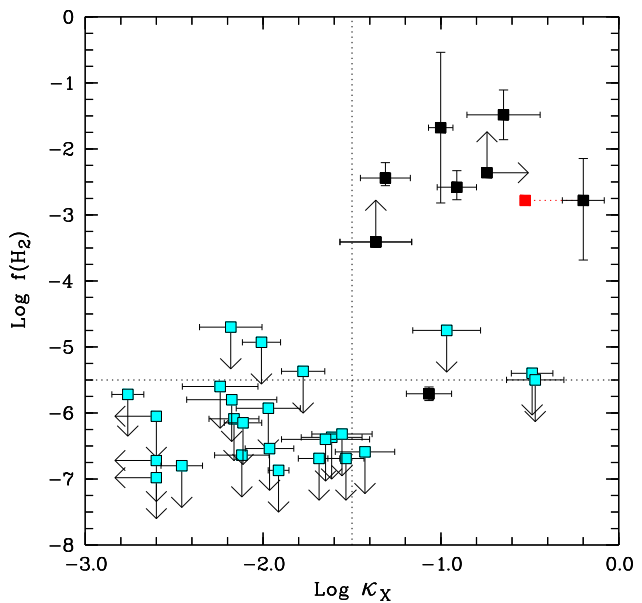,width=8.4cm,clip=,bbllx=42.pt,bblly=290.pt,bburx=554.pt,bbury=781.pt,angle=0.}}}
\caption[]{H$_2$ molecular fraction versus the amount of dust, or
dust-to-gas ratio $\kappa _{\rm X}=10^{\rm [X/H]}(1-10^{\rm [Fe/X]})$, with
either X$=$Zn, or S, or Si. Drawing conventions are the same as in
Fig.~\ref{figfdep}. A weak trend between the two quantities is
apparent; however, several orders of magnitude spread in $f($H$_2)$ for
a given dust content, among the detected cases, suggests that in addition
to the amount of dust the physical conditions of the gas (density,
temperature, UV flux) play an important role in governing the formation
of H$_2$ in DLA systems.}
\label{figfdust}
\end{figure}
%
%*****************************************************************************

The sharp transition from $\log f<-4$ to $\log f>-2$ at a column
density $\log N($H\,{\sc i}$)\approx 20.7$ observed for the Galactic lines of
sight (Savage et al. 1977) is not clearly present in the SMC, in the LMC nor
in sample SH$_2$. This transition is believed to occur at the point at which
the clouds achieve self-shielding. However, a characteristic H\,{\sc i}
column density can possibly correspond to some self-shielding scale only if
the parameters controlling the formation equilibrium of H$_2$ are
similar along different lines of sight, with very little scatter, which may
be the case in our Galaxy. Therefore, our results suggest that DLA
systems span a wide range of physical conditions. This is also suggested
by the large scatter seen in the metallicity measured in DLA systems at a
given redshift (see e.g. Prochaska \& Wolfe 2002). This conclusion is
strengthened by the fact that there is a clear dichotomy in $\log f$ values
whatever the H\,{\sc i} column density is, with very few measurements in
the range $-5<\log f<-3$. This bimodal distribution could be a consequence of
self-shielding, which would mean that, in DLA systems, self-shielding can
be achieved at any H\,{\sc i} column density probably because the UV radiation
field varies strongly from one system to the other. Alternatively, as shown
by Tumlinson et al. (2002) in the case of the Magellanic Clouds, most of
the observations can be explained by models where the formation rate of H$_2$
onto dust grains is reduced and the ionizing flux is enhanced relative to what
is observed in our Galaxy (see also the models by Shaye 2001). Basically, the
above characteristic H\,{\sc i} column density,
$\log N($H\,{\sc i}$)=20.7$, observed for Galactic clouds, would be located
beyond the observed range in the case of DLA systems (i.e.
$\log N($H\,{\sc i}$)\ga 21.8$).

Note that there are two sub-DLA systems, with $\log N($H\,{\sc i}$)<20.3$,
having detected H$_2$ molecules. These systems, at $z_{\rm abs}=1.968$ toward
Q\,0013$-$004 and $z_{\rm abs}=2.087$ toward Q\,1444$+$014, are amongst the
most metal (and dust) -rich absorbers of sample S1. They, together with the
DLA system at $z_{\rm abs}=2.338$ toward Q\,1232$+$082, have the highest
molecular fractions, and could be part of the population arising
in disc-like gas.

\subsection{The role of dust}

It can be seen on Fig.~\ref{figsample1} that the probability of detecting
H$_2$ is unity for systems with large depletion factors, [X/Fe$]\ge 0.7$. This
is a crude indication that the presence of dust is an important factor for
the formation of H$_2$ in DLA systems. It is known that H$_2$ formation is
mediated by either dust grains if the gas is cool and dense, or H$^-$ if it is
warm and dust-free. If the former process is dominant in DLA systems then
some correlation between H$_2$ molecular fraction and amount of dust is
expected. However, in the case of DLA systems it is difficult to quantify the
amount of dust that is present. We first plot in Fig.~\ref{figfdep}
molecular fractions versus depletions (probably onto dust grains) as estimated
from the abundances of Zn relative to Fe, or S when Zn is not detected, or
Si when Zn is not detected and sulfur lines are blended. A trend between
both quantities is present at the $3.6\sigma$ significance level using a
Kendall rank correlation test taking into account upper and lower limits on
$\log f$. Note however that there is a two orders of magnitude scatter in
$\log f$ for a given depletion factor so that there is basically
no correlation between $\log f$ and [X/Fe] if only detections are
considered (correlation significant at the $\la 1\sigma$ level only). The
above trend is therefore a consequence of H$_2$ being detected in none of
the systems with [X/Fe$]<0.4$ except one (at $z_{\rm abs}=2.595$
toward Q\,0405$-$443). The corresponding systems have a very low dust content.
A depletion factor [Zn/Fe$]\sim 0.4$ corresponds to a very small $E(B-V)$
($\sim 0.01$) for physical conditions similar to the ones prevailing in
our Galaxy.
We have seen that the depletion factor is correlated with the metallicity
in DLA systems. There is therefore some correlation between the
depletion factor and the amount of dust. However, the scatter is large.
This is why it can be, in general, misleading to use only the depletion
factor as an indicator of the dust content of the absorbing gas.

\begin{figure}
\flushleft{\vbox{
\psfig{figure=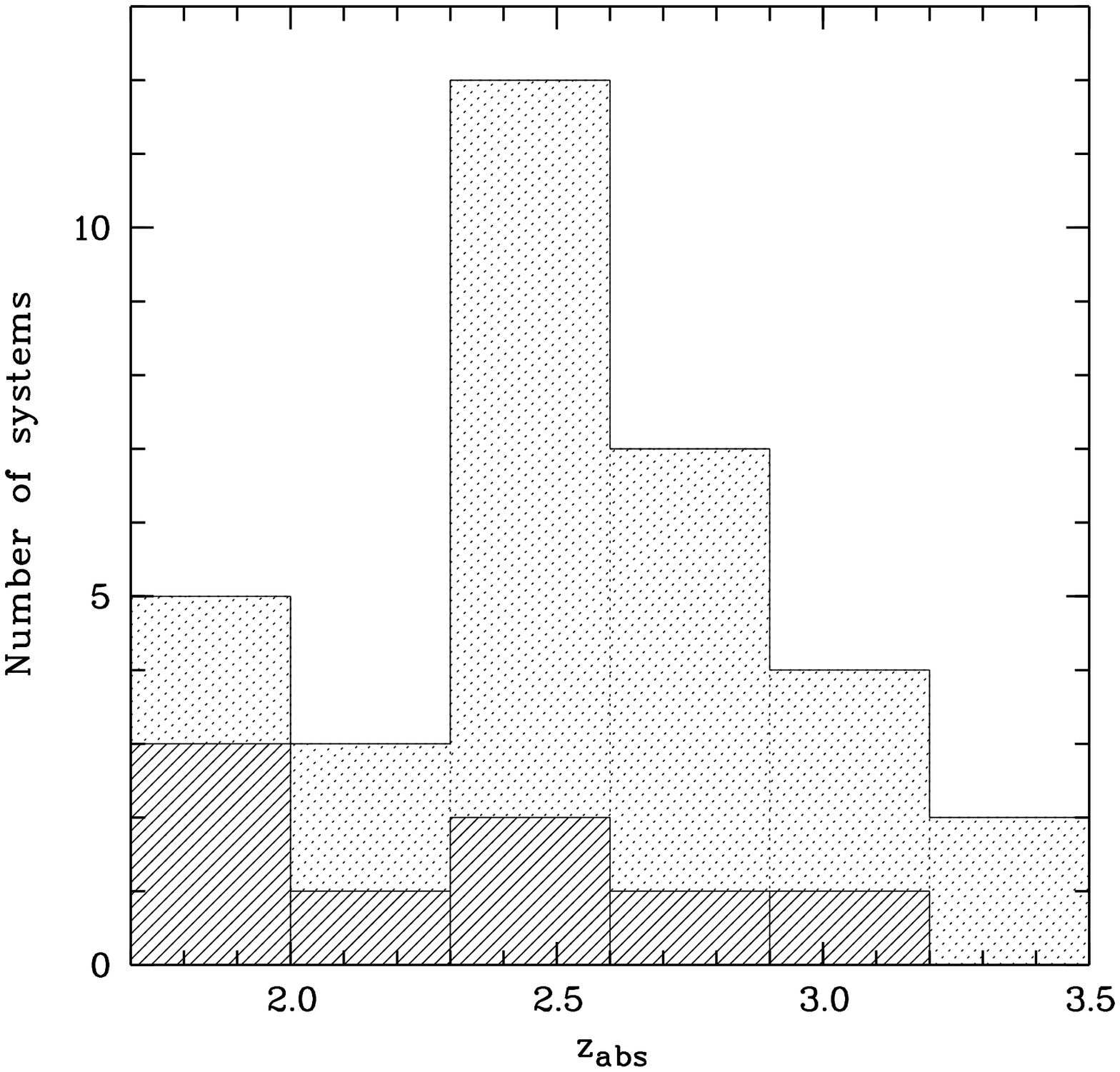,width=8.4cm,clip=,bbllx=46.pt,bblly=291.pt,bburx=554.pt,bbury=778.pt,angle=0.}}}
\vspace{-0.82cm}\hspace{-0.165cm}
\centerline{\hbox{
\psfig{figure=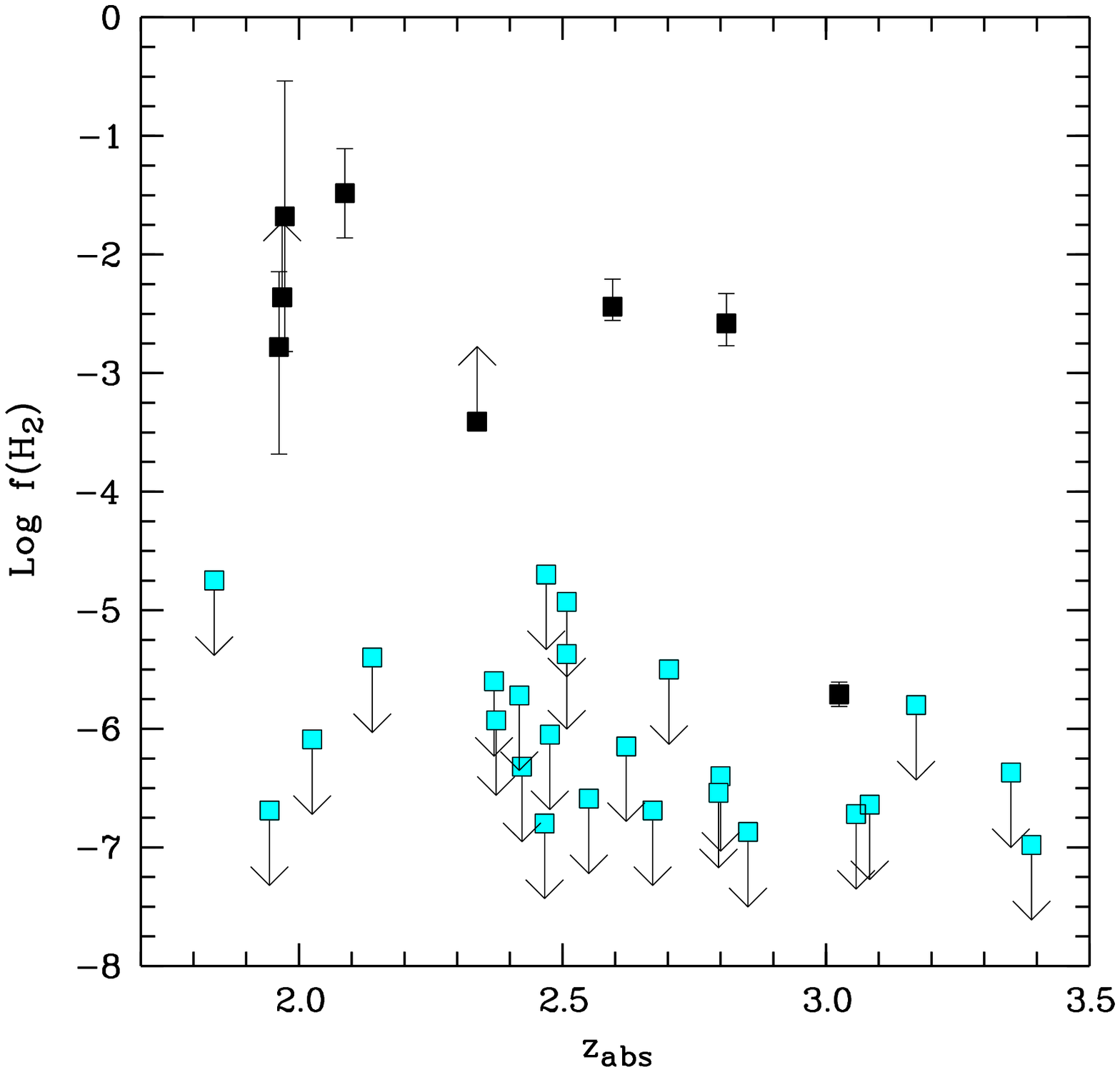,width=8.4cm,clip=,bbllx=46.pt,bblly=291.pt,bburx=554.pt,bbury=771.pt,angle=0.}}}
\caption[]{{\sl Upper panel}: redshift distribution of the DLA/sub-DLA systems
in our sample. The sub-sample of H$_2$-detected systems is indicated by
the hashed histogram. {\sl Lower panel}: H$_2$ molecular fraction versus
redshift. Detections are shown by dark squares and upper limits by shaded
ones. Although the statistics are based on small numbers, the probability of
finding molecules at $z_{\rm abs}\sim 2$ is highest. This probably
reveals that the amount of dust in DLA systems increases with time.}
\label{figfz}
\end{figure}

A better indicator of the dust content is the dust-to-gas
ratio: $\kappa _{\rm X}=10^{\rm [X/H]}(1-10^{\rm [Fe/X]})$, where X stands for
a reference element which is little or unaffected by dust depletion effects.
In Fig.~\ref{figfdust}, we plot the molecular fraction versus the amount of
dust as measured by the dust-to-gas ratio. It can be seen that, as expected,
low dust content, $\log\kappa _{\rm X}<-1.5$ (e.g. [X/H$]<-1.3$
and [X/Fe$]<0.5$), implies low molecular fraction, $\log f<-5$, while
larger molecular fraction, $\log f\ga -4$, is only found for higher dust
content, $\log\kappa _{\rm X}>-1.5$. However, large amounts of dust do not
always imply large molecular fractions. There are a few non-detections with
$\log\kappa _{\rm X}>-1.5$ and $\log f<-5$. In particular, the two
systems with $\log\kappa _{\rm X}\simeq -0.5$ and $\log f<-5$
are sub-DLA systems (see also Fig.~\ref{figsample2}, upper plot,
at [X/H$]\simeq -0.3$). Therefore, in these systems the lack of molecules
could be the consequence of enhanced radiation field below the Lyman limit.

Another important factor governing the presence of H$_2$ molecules appears to
be the local physical conditions of the absorbing gas. Relatively
large molecular fractions, $\log f>-4$, are found in gas having large
particle densities, $n_{\rm H}>20$ cm$^{-3}$, and low temperatures, $T<300$ K
(see Petitjean et al. 2002; Ledoux et al. 2002b).
One should keep in mind that the formation rate of H$_2$ goes linearly
with the density of dust grains while it goes as the second power of
the H\,{\sc i} density. Therefore, it is natural that, even though the
presence of dust is an important factor for the formation of molecules,
local physical conditions such as gas density, temperature and local UV field
play a major role in governing the molecular fraction of a given cloud in
DLA systems.

Note that the above discussion is based on the determination of mean
metallicities in DLA systems. In most cases, this means that the metallicity
is calculated over some velocity range and, therefore, over several
components. However, we have shown that the depletion in the components where
H$_2$ is detected can be much larger than the mean depletion. As an example, a
large depletion factor has been observed in a well-defined,
weak H$_2$ component in the DLA system at $z_{\rm abs}=1.97$
toward Q\,0013$-$004 (Petitjean et al. 2002). This kind of component can
usually be revealed by the corresponding, narrow C\,{\sc i} absorption
line. However, for other species the component is most often blended with
other components in the rest of the profile. This means that absorption
lines due to cold gas, i.e. related to C\,{\sc i} components, can be hidden in
the absorption line profile produced by a more pervasive medium. This implies
that absolute metallicities are not accurately determined in the cold
gas. This fact is probably not important for volatile elements as for them the
dominant components actually produce the strongest absorption lines, but could
be of importance for Fe co-production elements which can be depleted onto dust
grains. Consequently, absorption lines from the latter elements can be
weak and lost in the profiles. Therefore, we could underestimate the depletion
factors in these components.

\subsection{Evolution with redshift}

In the previous Section, we have given two new arguments in favor of the
presence of dust in DLA systems. First, there is a correlation between
metallicity and depletion factor and, second, molecular hydrogen is observed
in the systems having the highest metallicities and the largest
depletion factors. In a simple picture of star-formation history, one can
imagine that the gas is slowly enriched by on-going star formation
and, therefore, metallicity and dust content should increase together
with time. In Fig.~\ref{figfz}, we plot the total number of DLA/sub-DLA
systems in our sample versus redshift along with the number of systems
where H$_2$ is detected. Although the statistics are based on small
numbers, the probability of finding molecules in the
$z_{\rm abs}\sim 2$ systems is highest. This probably indicates that
the amount of dust in DLA systems increases with time. Although the
increase of Zn metallicity with decreasing redshift is not fully
established (e.g. Pettini et al. 1999), there is some evidence for such an
increase at $z_{\rm abs}<3.5$ (Prochaska \& Wolfe 2002;
Kulkarni \& Fall 2002). Consequently, the possible lack of increase with
time of the mean Fe metallicity in DLA
systems (e.g. Prochaska \& Wolfe 2002) could be related to the fact that
the intrinsic increase in metallicity is hidden by larger depletion onto dust
grains at low and intermediate redshifts (see Ledoux et al. 2002a).

\section{Conclusions}\label{con}

Although the presence of dust in DLA systems was claimed very early
(Pei, Fall \& Bechtold 1991), the issue of whether the abundance
pattern observed in DLA
systems reflects depletion of refractory elements onto dust grains or
nucleosynthesis effects has remained controversial (see e.g. Lu et al. 1996;
Pettini et al. 1997). Recently, several studies have shown that the two
effects, dust depletion and peculiar nucleosynthesis history, should
be invoked altogether to explain the
observations (Vladilo 1998; Prochaska \& Wolfe 2002; Ledoux et al. 2002a).
In any case, all these studies conclude that the dust content of DLA systems
is usually small. However, it is possible that the current sample of
DLA systems is biased against high-metallicity and dusty absorbers.
Indeed, Boiss\'e et al. (1998) have noticed that there is a lack of absorbers
with both a large $N($H\,{\sc i}$)$ and a high metallicity. If
this dust-induced bias exists, however, it probably cannot lead
to underestimate the H\,{\sc i} mass in DLA systems by more than a factor
of two (Ellison et al. 2001).

In the course of a survey of 24 DLA and 9 sub-DLA systems, we have
confirmed four previous detections of molecular hydrogen and made three
new ones. The bulk of the DLA population is found to have low metallicities,
[X/H$]<-1.0$, together with small depletion factors, [X/Fe$]<0.5$. Although
molecules can be found along lines of sight with small depletion factors
([Zn/Fe$]=0.3$), they are generally found in the most metal-rich
systems, [X/H$]>-1.0$, with large depletion factors, [X/Fe$]>0.5$. This
clearly demonstrates the existence of dust in at least 20 percent of the DLA
systems. Moreover, we have found very large depletion factors in
two sub-systems where H$_2$ is detected, at $z_{\rm abs}=1.96822$
toward Q\,0013$-$004 (Petitjean et al. 2002), where [Zn/Fe$]=1.6$, and
at $z_{\rm abs}=2.08692$ toward Q\,1444$+$014 (see Sect.~\ref{q1444}),
where [Zn/Fe$]=1.4$. This shows that dust depletion can be quite large in
some of the DLA systems. The extinction is small in these cases
however, probably because of relatively small H\,{\sc i} column densities
(e.g. $\log N($H\,{\sc i}$)<19.5$ in the $z_{\rm abs}=1.96822$ system toward
Q\,0013$-$004). In addition, there is a correlation between metallicity
and depletion factor in all DLA samples (large depletion factors are seen
in systems with high metallicities). This trend has been noticed before by
Ledoux et al. (2002a). However, it was unclear whether this trend is
really due to dust depletion. Prochaska \& Wolfe (2002) indeed argued that
such a correlation may arise from a significant dispersion in the production
of zinc. However, the correlation is present in both our sample and also the
global population of DLA systems. Moreover, it is clear that
molecular hydrogen is mostly detected at the highest metallicities and
the largest depletion factors. This strongly favors the hypothesis that
the correlation between metallicity and depletion factor is actually a
consequence of dust depletion effects. Not only dust could be present in
most of the DLA systems, but also the possibility that the population of DLA
systems is biased against the presence of systems having both a high
metallicity and a large dust content still needs to be considered seriously.

One of the striking results of the survey is that for most of the DLA systems
the molecular fraction is very small ($\log f<-6$). This is much smaller than
what is observed in the disc of the Galaxy at similar H\,{\sc i} column
densities (see Savage et al. 1977). However, the situation is comparable to
what is observed along lines of sight in the LMC and SMC
(Tumlinson et al. 2002). Indeed, similar upper limits on the
molecular fraction have been measured for a large number of lines of sight in
the LMC. Moreover, the mean molecular fraction in DLA systems where H$_2$ is
detected is of the order of 0.01, about the same as in the LMC and SMC
and about a factor of ten smaller than in the disc of the Galaxy. This is
probably a consequence of reduced formation rate of H$_2$ onto dust
grains because the temperature of the gas is
high (Petitjean et al. 2000). Indeed, in the framework of models of
cool clouds it is difficult to explain the very low values of the upper
limits ($\log f\la -7$) achieved for a handful of DLA systems in our sample
(see fig.~7 of Liszt 2002). This can also be due to an intense ambient UV flux
(see models by Tumlinson et al. 2002), but this is unclear and should be
investigated in detail by analysis of the systems where H$_2$ is detected (see
Srianand \& Petitjean 1998; Petitjean et al. 2002; Ledoux et al. 2002b).

Star-formation activity is probably not intense in the close vicinity of
most of the systems. M\o ller et al. (2002) found that a few DLA systems
are associated with Lyman-break galaxies. However, from spectroscopy of 11 out
of 22 candidates in 6 fields, targeting 8 DLA systems, they identified
only one galaxy counterpart. This means that, even if DLA systems can
be associated with Lyman-break galaxies, they must have most of the
time moderate star-formation activity with luminosities $L<L^*$
(see e.g. Fynbo, M\o ller \& Thomsen 2001). In addition, in the framework
of current
models of structure formation, DLA systems preferentially sample the outer
regions of galaxies at the faint end of the galaxy luminosity function
(Haehnelt et al. 2000). One way of explaining the small molecular fractions
observed in DLA systems is that the associated emitting object is faint
and has moderate star-formation activity, but that the gas giving rise to the
DLA absorption line is located inside the regions where star formation occurs.
The probably large turbulence induced by on-going star formation may then
help explain the required large covering factors associated with such
regions. This would imply that star formation is gentle and diffuse in
the systems. The diffuse UV flux can be estimated from the excitation of
molecular hydrogen. This will be done in a forthcoming paper. Additional deep
imaging in the vicinity of the QSO lines of sight is also needed to unveil the
nature of star-formation activity in DLA systems.

\section*{acknowledgements}
It is a pleasure to thank our colleagues from ESO Paranal observatory,
and in particular A. Kaufer, for their patient, thorough and efficient help
at the VLT and during data reduction. We also warmly thank J. Fynbo and
an anonymous referee for useful comments on the paper and J. Smoker for a
careful reading of it. CL acknowledges support from an
ESO post-doctoral fellowship during the time the survey was carried out.
PPJ thanks ESO for an invitation to stay at its headquarters in Chile where
part of this work was completed. The project was supported by the
European Community Research and Training Network: `The Physics of the
Intergalactic Medium'. RS and PPJ acknowledge support from the Indo-French
Centre for the Promotion of Advanced Research (Centre Franco-Indien pour la
Promotion de la Recherche Avanc\'ee) under contract No. 1710-1.

\label{lastpage}

\end{document}